\documentclass[aps,prd,twocolumn,groupedaddress,showpacs]{revtex4}

\usepackage{graphicx} \usepackage{color} \usepackage{natbib}

\usepackage{amsmath,amssymb,amsfonts}
\usepackage{epsf}
\usepackage{epsfig}
\usepackage{longtable}
\begin{document}


\title{Discriminating strange star mergers from neutron star mergers\\
 by gravitational-wave measurements} 


\author{A.~Bauswein, R.~Oechslin, H.-T.~Janka}
\affiliation{Max-Planck-Institut f\"ur
  Astrophysik, Karl-Schwarzschild-Str.~1, D-85748 Garching, Germany}


\date{\today}

\begin{abstract}
We perform three-dimensional relativistic hydrodynamical simulations of the coalescence of strange stars and explore the possibility to decide 
on the strange matter hypothesis by means of gravitational-wave measurements. Self-binding of strange quark matter and the generally more compact stars yield features that clearly distinguish strange star from neutron star mergers, e.g. hampering tidal disruption during the plunge of quark stars. Furthermore, instead of forming dilute halo structures around the remnant as in the case of neutron star mergers, the coalescence of strange stars results in a differentially rotating hypermassive object with a sharp surface layer surrounded by a geometrically thin, clumpy high-density strange quark matter disk. We also investigate the importance of including nonzero temperature equations of state in neutron star and strange star merger simulations. In both cases we find a crucial sensitivity of the dynamics and outcome of the coalescence to thermal effects, e.g. the outer remnant structure and the delay time of the dense remnant core to black hole collapse depend on the inclusion of nonzero temperature effects. For comparing and classifying the gravitational-wave signals, we use a number of characteristic quantities like the maximum frequency during inspiral or the dominant frequency of oscillations of the postmerger remnant. In general, these frequencies are higher for strange star mergers. Only for particular choices of the equation of state the frequencies of neutron star and strange star mergers are similar. In such cases additional features of the gravitational-wave luminosity spectrum like the ratio of energy emitted during the inspiral phase to the energy radiated away in the postmerger stage may help to discriminate coalescence events of the different types. If such characteristic quantities could be extracted from gravitational-wave signals, for instance with the upcoming gravitational-wave detectors, a decision on the strange matter hypothesis and the existence of strange stars should be possible.
\end{abstract}

\pacs{04.30Tv, 12.39Ba, 21.65Qr}

\maketitle

\section{Introduction}
The equation of state (EoS) of high-density matter and therefore the true nature of compact stars has been a mystery since the discovery of these objects. Besides the possibility of a neutron-proton composition more exotic phases have been proposed to appear in the cores of compact stars (see e.g. \cite{2007ASSL..326.....H} for a review, for early considerations of quark stars see~\cite{1969NCimL...2...13I,1970PThPh..44..291I}). The formulation of the strange matter hypothesis \cite{PhysRevD.4.1601,PhysRevD.30.272} introduced the possibility that compact stars are so-called strange stars (SSs) \cite{1986A&A...160..121H,1986ApJ...310..261A}. According to this hypothesis the absolute ground state of matter is formed by a quark phase consisting of up, down and strange quarks, and compact stars are self-bound objects composed of this kind of matter. It turns out that SSs are in many ways similar to neutron stars (NSs), e.g. in the mass range and compactness, and therefore they can be considered as an alternative explanation for compact stars \cite{1986A&A...160..121H,1986ApJ...310..261A,1996csnp.book.....G,2007ASSL..326.....H,2005PrPNP..54..193W}. Throughout this article the term compact star refers to either a NS or a SS.

Theoretically as well as observationally, the question whether the strange matter hypothesis is true and whether (at least some) NSs are actually SSs, is an open issue. Besides the determination of compact star parameters, several possibilities have been proposed to explore the observational consequences of the strange matter hypothesis, including the possibility that small lumps of this strange quark matter (SQM) with baryon numbers of about $10^2$ and more might be abundant in the cosmic ray flux (see e.g. \cite{2005PhRvD..71a4026M}). Experiments like AMS-02, which is planned to be installed on the International Space Station in 2010, have the potential to detect these so-called strangelets \cite{ams,2004JPhG...30S..51S}. Also, SQM might be detected with the LHC at CERN \cite{SchaffnerBielich:1996eh,Greiner:1987tg,Spieles:1996is}. For a review on these considerations and other observational implications of the strange matter hypothesis see \cite{2005PrPNP..54..193W,2006JPhG...32S.251F}.

With the current and upcoming GW detectors like LIGO \cite{:2007kva} and VIRGO \cite{Acernese:2006bj} and the prospect of a detection of signals from compact object binaries \cite{2004ApJ...601L.179K}, the question naturally arises if the signal of mergering NSs could be distinguished from SS mergers, especially because compact object binaries are considered to be among the most promising sources for these detectors.

The systematic investigation of the imprint of the EoS on the gravitational-wave (GW) signal is still in its infancy. Most studies consider simplified EoSs like polytropes and try to explore the chances to measure some general compact star properties like the stellar radius (e.g. \cite{1996PhRvD..54.7261Z,2009arXiv0901.3258R}). It is not clear whether a decision on the strange matter hypothesis could be made on the basis of such measurements, because the compactness of SSs can be in the same range as that of NSs. First attempts of exploring the consequences of microphysical EoSs including nonzero temperature effects in a relativistic treatment, necessary for reliable results, have been made by \cite{2007PhRvL..99l1102O}. Also fully relativistic studies have been conducted in which temperature corrections were approximated by an ideal gas component added to a zero-temperature microphysical EoS (e.g. \cite{2006PhRvD..73f4027S} and preceding works).

Strange stars as sources of GWs have been considered in a rotational equilibrium approach investigating the (final phase) of the inspiral of SS binaries \cite{PhysRevD.71.064012,GondekRosinska:2008nf}. Quasiequilibrium orbits were constructed to determine the innermost stable circular orbit (ISCO). It was found that the frequency of the GWs at the ISCO depends on the compactness of the SSs. The orbital evolution and the associated GW emission of SS-black hole binaries has been estimated within a semianalytic model in \cite{2004JPhG...30.1279P}. Moreover, the signals from various instabilities of rotating SSs have been worked
out in \cite{2000PhRvL..85...10M,2002MNRAS.337.1224A,GondekRosinska:2003iy} (see also references therein). In \cite{2008PhRvL.101r1102F} it was found that the frequencies
of the g-mode in newly born SSs are significantly lower than
those of NSs. Differences of the fundamental pressure modes are discussed in \cite{2007GReGr..39.1323B} where a discrimination was found only to be possible if additional information like the mass was available.
Nevertheless it is not clear whether these discriminating features are relevant as the corresponding GW signals might be too weak for measurements.

In this article we examine the GW signals emitted by merging SSs. We explore two different EoSs representing different possibilities of SQM properties and consider the inspiral phase, the final plunge, and the postmerger stage. We compare the GW characteristics of SS coalescence with those of ordinary NSs and discuss if GW observations could be decisive on the strange matter hypothesis. For that we conduct three-dimensional relativistic hydrodynamical simulations that allow us to follow the evolution of the whole merging process and to extract observational signatures. To our knowledge this is the first study considering these events in dynamical simulations; only results for mergers of SS-black hole systems have been so far reported in \cite{2002MNRAS.335L..29K}, where Newtonian hydrodynamics were used and the black hole (BH) was modeled by a pseudorelativistic potential \cite{1980A&A....88...23P}.

The paper is structured as follows: Sect.~\ref{sec:num} will introduce the underlying model and numerical methods. In Sect.~\ref{sec:eos} the properties of the EoSs used in this study will be described. The simulations and some general aspects will be discussed in Sect.~\ref{sec:sim}. In Sect.~\ref{sec:GW} the GW signals will be presented and in Sect.~\ref{sec:con} we will close with a summary and our conclusions.

\section{Numerical methods}\label{sec:num}

For this study we used a version of the general relativistic smooth particle hydrodynamics code described in \cite{2004MNRAS.349.1469O,2007A&A...467..395O} and we refer the reader to these publications for details. As in grid-based Eulerian relativistic hydrodynamics schemes we introduce a set of so-called conserved quantities $(\rho^{*},\hat{u}_i, \tau)$, the conserved rest mass density, the conserved specific momentum and the conserved energy density, which relate to the primitive quantities $(\rho,v_i,\epsilon)$, the rest mass density, the coordinate velocity and the specific internal energy, via 
\begin{eqnarray}
\rho^* &=&\rho \alpha u^0 \psi^6,\label{eq:rhostar}\\
\hat{u}_i &=&hu_i=h(v^i+\beta^i)\psi^4 u^0,\\
\tau &=&h\Gamma-\frac{P}{\rho \Gamma}-\sqrt{1+\frac{\hat{u}_i \hat{u}_j \delta^{ij}}{\psi^4}}.\label{eq:energy}
\end{eqnarray}
Here the Lorentz factor is defined by $\Gamma=\alpha u^0=(1+\gamma^{\mathrm{ij}}u_i
u_j)^{1/2}$, $u^0$ and $u_i$ are the time and space components of the
eigenvelocity, respectively, $h=1+P/\rho +\epsilon$ represents the specific
relativistic enthalpy, and the metric potentials $\alpha$, $\psi$ and $\beta^i$ are introduced below.

The evolution of these conserved quantities can be expressed in a Lagrangian manner, e.g. comoving with the fluid (see \cite{2004MNRAS.349.1469O,2007A&A...467..395O}). Having at hand this Lagrangian formulation of relativistic hydrodynamics the discretization according to the Smoothed Particle Hydrodynamics (SPH) method can be done in a straightforward manner (see e.g. \cite{2000ApJ...531.1053S,2004MNRAS.349.1469O}) by defining a smoothing operator
\begin{equation}
\label{eq:int}
\langle A(r)\rangle=\int A(r') W(|r-r'|;h)d^3r'
\end{equation}
for any quantity $A(r)$. The analytically given normalized, continuous, differentiable kernel function $W(r;h)$ has compact support and is characterized by the smoothing length $h$ \cite{1985A&A...149..135M}. Representing the fluid by a sample of particles with fixed rest masses $m_a$ the above integral can be approximated by a summation
\begin{equation}
\langle A(r)\rangle\approx \sum_a A(r_a) \frac{m_a}{\rho^*_a}W(|r-r_a|;h)
\end{equation}
over neighboring particles, where $\rho^*_a$ is the conserved rest mass density of particle ``a'' located at $r_a$. Rewriting the evolution equations in this form has the advantage that partial derivatives can be expressed by means of an integration by parts in equation (\ref{eq:int}) as derivatives of $W$, which are analytically given. Therefore one saves the computationally expensive evaluations of partial derivatives.

In our code the resulting system of ordinary differential equations (see~\cite{2007A&A...467..395O}) is solved with a fourth order Runge-Kutta method. In every timestep one recovers the primitive quantities $(\rho,v_i,\epsilon)$ from the evolved quantities $(\rho^{*},\hat{u}_i, \tau)$ by solving numerically the system \eqref{eq:rhostar}-\eqref{eq:energy}, where the temperature $T$ is determined from the specific internal energy $\epsilon$ by means of the EoS. The EoS $P(\rho,Y_e,T)$ finally closes the system of hydrodynamical equations and will be discussed in Sect.~\ref{sec:eos}. The initial electron fraction of the cold stars is defined by the condition of beta-equilibrium and is assumed to be advected with the fluid ($\frac{dY_e}{dt}=0$) because the timescale of lepton number transport (at least for the bulk of the stellar matter in the interior, where the GW signal is determined) is long compared to the dynamical timescale. This, however, does not hold for the matter shed off the surface of the merging stars, where lepton number and energy loss may be important.

In contrast to standard Newtonian SPH, gravity cannot be implemented within the particle picture (e.g. with a tree method). Instead, the Einstein field equations have to be solved. For this we employ the conformal flatness approximation, which assumes that the metric can be written as $ds^2=(-\alpha^2+\beta_i \beta^i)dt^2+2\beta_i dx^i dt+\gamma_{\mathrm{ij}}dx^i dx^j$ with $\gamma_{\mathrm{ij}}=\psi^4 \delta_{\mathrm{ij}}$. Here $\alpha$ is the lapse function, $\beta^i$ is the shift vector and $\gamma_{\mathrm{ij}}$ is the spatial part of the metric approximated by the conformal factor $\psi$ and the Kronecker delta $\delta_{\mathrm{ij}}$ \cite{1980grg..conf...23I,1996PhRvD..54.1317W}. Within this approach the Einstein field equations reduce to a set of five coupled nonlinear elliptic partial differential equations with noncompact source terms. These equations are discretized on a grid covering a domain around the binary, respectively the postmerger remnant, and are solved with a full multigrid method.

Since GWs are neglected by this ansatz for the metric, a scheme simulating the backreaction of the GW emission has to be implemented in order to account for the loss of energy and angular momentum carried away by the gravitational radiation, which drives the binary to inspiral and to finally merge. Following the ideas of \cite{1990MNRAS.242..289B,2003PhRvD..68h4001F} this is done in a post-Newtonian framework. For details of the implementation of the GW backreaction see \cite{2007A&A...467..395O}, where the code we use is described in all parts, including the treatment of shocks.

Magnetic fields are not included in our implementation. The effects of magnetic fields smaller than $10^{15} G$ on the merger dynamics are negligible on shorter than secular timescales \cite{2008PhRvD..78b4012L,2009arXiv0901.2722G}.

\subsection{Initial data}
Prior to the actual merger simulations the binaries are evolved without the effects of the GW backreaction in order to find the corresponding quasiequilibrium state of the configuration. This is achieved by the relaxation of the particle distribution, which is slightly off equilibrium after setting up the two individually hydrostatic stars in a close orbit at the beginning.

The initial velocity profile is assumed to be irrotational because the shear viscosity of SQM is comparable to that of nuclear matter \cite{Haensel:1991pi}. Therefore the time during inspiral is not sufficient to yield tidally locked systems \cite{1992ApJ...400..175B}. Also because of time arguments it is reasonable to start with initially cold SSs in beta-equilibrium.

\section{Equation of state}
\label{sec:eos}
The EoSs of SQM are derived within the MIT bag model \cite{PhysRevD.9.3471,Farhi:1984qu}. Initially developed to describe nucleons, the application to bulk quark matter is rather simple. In a certain volume, the bag, the quarks are considered to be free Fermi gases. The masses of the up and down quarks can be neglected, while for the strange quark mass a value of $m_S=100~\mathrm{MeV}$ is adopted \cite{Eidelman:2004wy}. The confinement of QCD is simulated by a nonzero pressure of the vacuum outside onto the bag given by the bag constant $B$. Within the MIT bag model the pressure $P$, the energy density $e$ and the baryon density $\rho$ read \cite{1996csnp.book.....G}
\begin{eqnarray}
P&=&\sum_f\frac{1}{3} \frac{\gamma_{{f}}}{2\pi^2} \int^{\infty}_0 dk\, k^3\frac{\partial\epsilon_{{f}}(k)}{\partial k}[ n(k,\mu_{{f}})\nonumber\\ &&+n(k,-\mu_{{f}})] -B,\\
e&=& \sum_{{f}}\frac{\gamma_{{f}}}{2\pi^2} \int^{\infty}_0 dk\,k^2\epsilon_{{f}}(k)[ n(k,\mu_{{f}})\nonumber\\
&&+n(k,-\mu_{{f}})] +B,\\
\rho&=&\sum_{{f}}\frac{1}{3} \frac{\gamma_{{f}}}{2\pi^2} \int^{\infty}_0dk\,k^2\left[ n(k,\mu_{{f}}) -n(k,-\mu_{{f}})\right],
\end{eqnarray}
where $\epsilon_{{f}}(k)=(m_{{f}}^2+k^2)^{1/2}$ and the Fermi-Dirac distribution function $n(k,\mu_{{f}})=\left( \exp \left( \left[ \epsilon_{{f}}(k)-\mu_{{f}} \right]/T \right) +1 \right)^{-1}$ for a temperature $T$ is used. The chemical potentials of the different quark flavors are denoted by $\mu_{f}$. Note that we set $c=\hbar=1$. The sum extends over all quark flavors with the degree of freedom $\gamma_{{f}}=2\times 3$ for two spin states and three color states. These equations can be solved analytically in two limits: for massive quarks and zero temperature, or alternatively for massless quarks with nonzero temperature \cite{1996csnp.book.....G}. However, in this study a massive strange quark is assumed and thermal effects are supposed to play a role during the merging process of SSs and therefore the above set of equations needs to be solved numerically \cite{Giuseppe}. The conservation of baryon charge and electric charge and the condition of electric charge neutrality lead to a set of constraints for the chemical potentials under which the above equations are solved. These constraints also determine the chemical potential of the electrons, which are treated independently as an ideal Fermi gas. For details we refer to \cite{1996csnp.book.....G,1999paln.conf.....W}.

The bag constant is a free but constrainted parameter of the model. Because a spontaneous conversion of nuclei into SQM is not observed, the lower limit is given by $B=57~\mathrm{MeV/fm^3}$. In order to obtain an energy per baryon lower than the one of ordinary nuclear matter the upper bound of $B$ equals $84~\mathrm{MeV/fm^3}$. The range of $B$ can slightly differ for other choices of $m_S$ or if one includes interactions between the quarks. These limits for $B$ correspond to an energy per baryon lower than that of nuclear matter ($E/A=930~\mathrm{MeV}$) and therefore they account for absolutely stable quark matter, i.e. the true ground state of matter. In this study we use the constant values $B=60 ~\mathrm{MeV/fm^3}$ ($E/A=860 ~\mathrm{MeV}$) and $B=80 ~\mathrm{MeV/fm^3}$ ($E/A=921 ~\mathrm{MeV}$) and refer to these EoSs as MIT60 and MIT80. Because of these choices of $B$ the EoSs roughly represent the two extreme cases of the underlying microphysical model. Potential extensions of the model could for instance account for the effects of color superconductivity or interactions between the quarks. In order to have a clear parameter dependence we omitted the inclusion of such corrections, which anyway could be absorbed in an ``effective bag constant'' \cite{Fraga:2001id}.

In principle SSs can carry a thin nuclear crust whose maximum density cannot exceed the neutron drip density \cite{1986ApJ...310..261A}. When free neutrons form they settle gravitationally and are converted into SQM when they come in contact with the quark phase. For this reason the mass of the crust is only of the order of $10^{-5}~M_{\odot}$. Therefore we neglect it in our simulations, because it is not of importance for the dynamics of the system.

Since we want to compare our findings for SS mergers with NS coalescence, we also performed simulations with EoSs describing NS matter. To this end we employed the same nuclear EoS as used in \cite{2007A&A...467..395O}, namely those of Refs. \cite{1998NuPhA.637..435S,1991NuPhA.535..331L} (see there for more details) and abbreviate them in the following by Shen and LS, respectively. These models also span a considerable range of variation for NS matter in the sense that LS yields fairly compact stars while Shen results in significantly less compact objects (Fig.~\ref{fig:MR}).

\begin{figure}
\includegraphics[width=8.9cm]{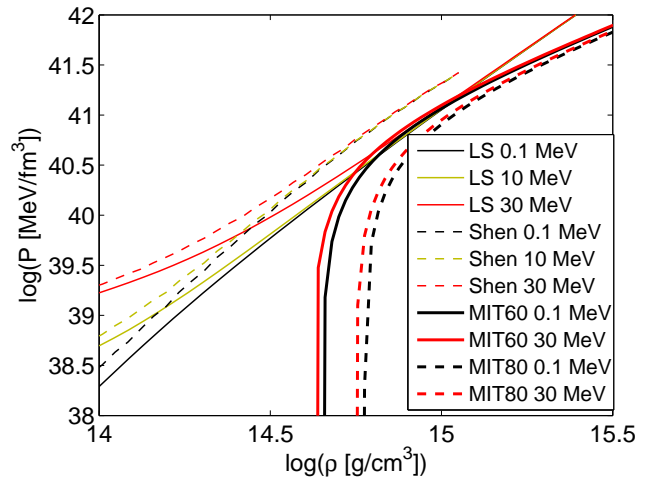}
\caption{\label{fig:Prho} Pressure as a function of rest-mass density $\rho=m_u n_B$ for the MIT60, MIT80, LS and Shen EoSs, and for different chosen temperatures. The EoSs for SQM can be easily recognized by their rapid decrease of the pressure below a certain density. The electron fraction for the nuclear EoSs is set to 0.05 and for the quark matter EoSs to $2.5\cdot 10^{-5}$.}
\end{figure}

\begin{figure}
\includegraphics[width=8.9cm]{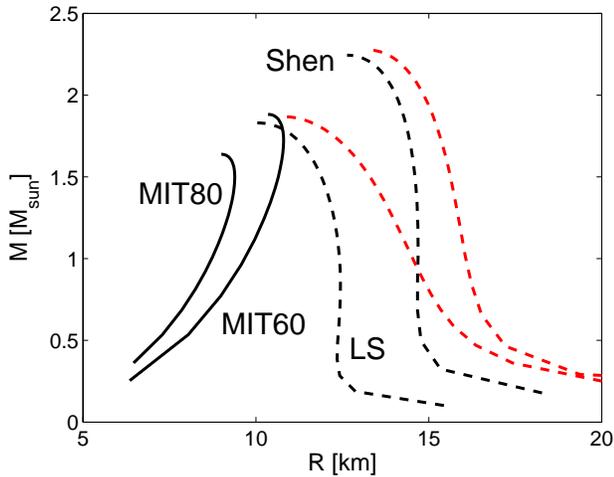}
\caption{\label{fig:MR}The stable branches of the mass-radius relations for compact stars with EoSs MIT60, MIT80, LS and Shen. $M$ refers to the gravitational mass in isolation and $R$ is the radius in Schwarzschild coordinates. The black curves display the mass-radius relations for $T=0$, whereas the red curves show them for isentropic stars with an entropy per nucleon of 1.1 $k_B$ for the Shen EoS (right curve) and 1.5 $k_B$ for the LS EoS (left curve). For the SQM EoSs the differences due to thermal effects are significantly less than 1\% even for the chosen entropy of 3.2 $k_B$ in the cases of the MIT60 EoS and 3.1 $k_B$ for the MIT80 EoS (see text for an explanation of the choices of the entropy values).}
\end{figure}

Figure~\ref{fig:Prho} shows for all EoSs used in this study the pressure as a function of the rest mass density $\rho=m_u n_B$ ($n_B$ is the baryon number density, $m_u$ the atomic unit mass) for representative temperatures of 0.1 MeV, 10 MeV and 30 MeV. For the EoSs of SQM one can clearly see the drop of the pressure towards zero at densities well above nuclear density ($2.7\cdot 10^{14}~\mathrm{g/cm^3}$). Moreover, in the case of SQM, except for a narrow range of densities close to this minimum density, the temperature has only a minor influence on the pressure. This is due to the fact that SQM is forming a degenerate Fermi gas of ultrarelativistic particles. Also for the nuclear EoSs thermal effects become more important the lower the density is. At a density of $10^{14}~\mathrm{g/cm^3}$ the increase of the temperature from 0.1 MeV to 30 MeV leads to a pressure increase of about one order of magnitude.

Solving the Tolman-Oppenheimer-Volkoff stellar structure equations for the EoSs introduced above, i.e. solving the general relativistic equations of hydrostatic equilibrium, one obtains the mass-radius relations of bare SSs in beta-equilibrium at $T=0$. Figure~\ref{fig:MR} shows the corresponding results with $M$ referring to the gravitational mass (Arnowitt-Deser-Misner mass) in isolation and $R$ being the radius in Schwarzschild coordinates. First of all, one recognizes the typical inverse mass-radius relation for the NS models, which is much different from that of SSs. The maximum mass of MIT60 and MIT80 is $1.88~M_{\odot}$ and $1.64~M_{\odot}$, respectively. Remarkably, the maximum mass of LS is $1.83~M_{\odot}$, thus comparable to the one of MIT60, while the Shen EoS supports stars with masses up to $2.24~M_{\odot}$. More massive nonrotating objects collapse to BHs. Note that the inclusion of rotation increases these maximum masses. Rigidly rotating stars more massive than these maximum nonrotating objects are called supermassive, while differential rotation, which increases the maximum mass further in comparison to uniform rotation, allows for so-called hypermassive stars exceeding the maximum mass of supermassive stars \cite{2000ApJ...528L..29B}. The stellar radii for LS and MIT60 in the high-mass range are similar and therefore the compactness $GM/c^2R$, too. In general, SSs are more compact than NSs, and MIT80 yields smaller stars than MIT60.

In order to illustrate the effects of nonzero temperature, we also plot the mass-radius relations for isentropic stars with nonzero entropy. The entropy is fixed by the value obtained when matter at about three times nuclear saturation density is chosen to have a temperature of 30 MeV, which are typical temperature and density values reached during a binary merger. This corresponds to an entropy per baryon of 1.1 $k_{B}$ for the Shen EoS and 1.5 $k_{B}$ for the LS EoS. The same criterion gives an entropy per baryon of 3.2 $k_B$ for the MIT60 EoS and 3.1 $k_B$ with the MIT80 EoS. One can see that thermal effects increase the maximum mass of NSs only slightly. However, the influence on the NS radius is bigger the smaller the mass is. Especially at lower densities thermal effects become important as mentioned above. Remarkably, doing the same for the SS EoSs yields mass-radius curves that are hardly distinguishable from those for $T=0$, which is also understandable from Fig.~\ref{fig:Prho}.

We stress the fact that SQM as a small nugget but also as an SS is self-bound by QCD interactions (and a disintegration of the constituents is energetically not favorable). As the pressure goes to zero the density at the surface of an SS (or a small nugget of SQM) adopts a value of the order of the nuclear saturation density (see Fig.~\ref{fig:Prho}). It implies that this type of matter cannot become arbitrarily dilute. This also means that for the binding of a SS gravity is not essential, it only compresses the star additionally. In contrast, an ordinary NS is solely bound by gravitation and would ``explode" without the effect of gravity. Pictorially spoken, SQM behaves hydrodynamically like a liquid with surface tension.

Finally, we note that the mass-radius relation is a crucial quantity determining the dynamics and the outcome of a merger (see below and \cite{2002PhRvL..89w1102F,2007A&A...467..395O}). In our SQM model the bag constant is the only parameter varied and thus affects entirely the mass-radius relation. Therefore our results can be interpolated for other choices of the bag constant between the two considered nearly extreme cases. Similarly, one can estimate how disregarded effects like quark interactions and color superconductivity might influence our results by inspecting their effects on the mass-radius relation (see also \cite{2008arXiv0812.4248B}). In the same way one can judge on the impact of other quark matter models beyond the standard MIT bag model. In Ref.~\cite{1995PhRvD..51.1989B}, for instance, running quark masses were considered, which affect the mass-radius relation such that SSs can have a higher maximum mass, which will lead to differences compared to our models with respect to the onset of BH formation.

\section{Simulations} \label{sec:sim}
\begin{figure}
\includegraphics[width=8.9cm]{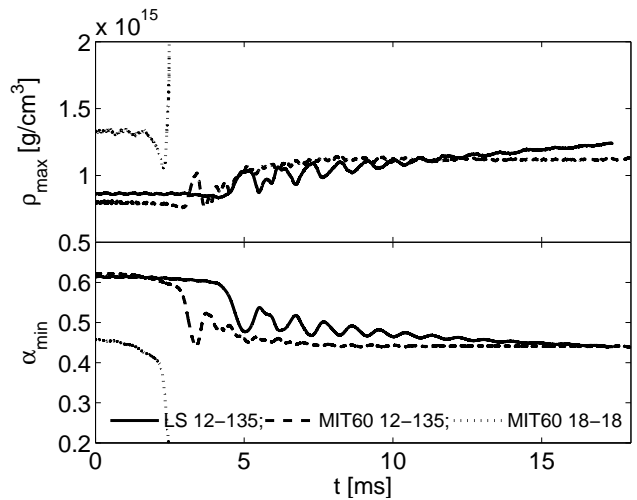}
\caption{\label{fig:rhomax} Time evolution of the maximal rest mass density (upper panel) and the minimal lapse function (lower panel). Note that we added a time shift of -4.5~ms to the 1.2$M_{\odot}$+1.35$M_{\odot}$ model and of -6~ms to the others to avoid overlapping curves.}
\end{figure}

The mergers of SSs fall into two categories similar to the case
of NS coalescence. For high total binary masses the
remnant of the merging stars cannot be supported against gravitational
collapse and a BH forms shortly after the stars have come into
contact. The forthcoming appearance of a BH is indicated by a steep increase of the central density $\rho_{\mathrm{max}}$ exceeding more than twice the maximum central density of nonrotating SSs accompanied by a steep decrease of the lapse function $\alpha$, both on a timescale smaller than the sonic time scale. This can be seen in Fig.~\ref{fig:rhomax} (dotted line) for a symmetric binary with two 1.8~$M_{\odot}$ stars described by the MIT60 EoS. On the other hand, a merged object with a sufficiently low total mass can
be temporarily stabilized mainly by differential rotation; a corresponding hypermassive object forms when the total mass exceeds the maximum mass of supermassive stars \cite{2000ApJ...528L..29B}. This object will also collapse to a BH after angular momentum redistribution.

\begin{figure*}[t]
\epsfxsize=3.4in
\leavevmode
\epsffile{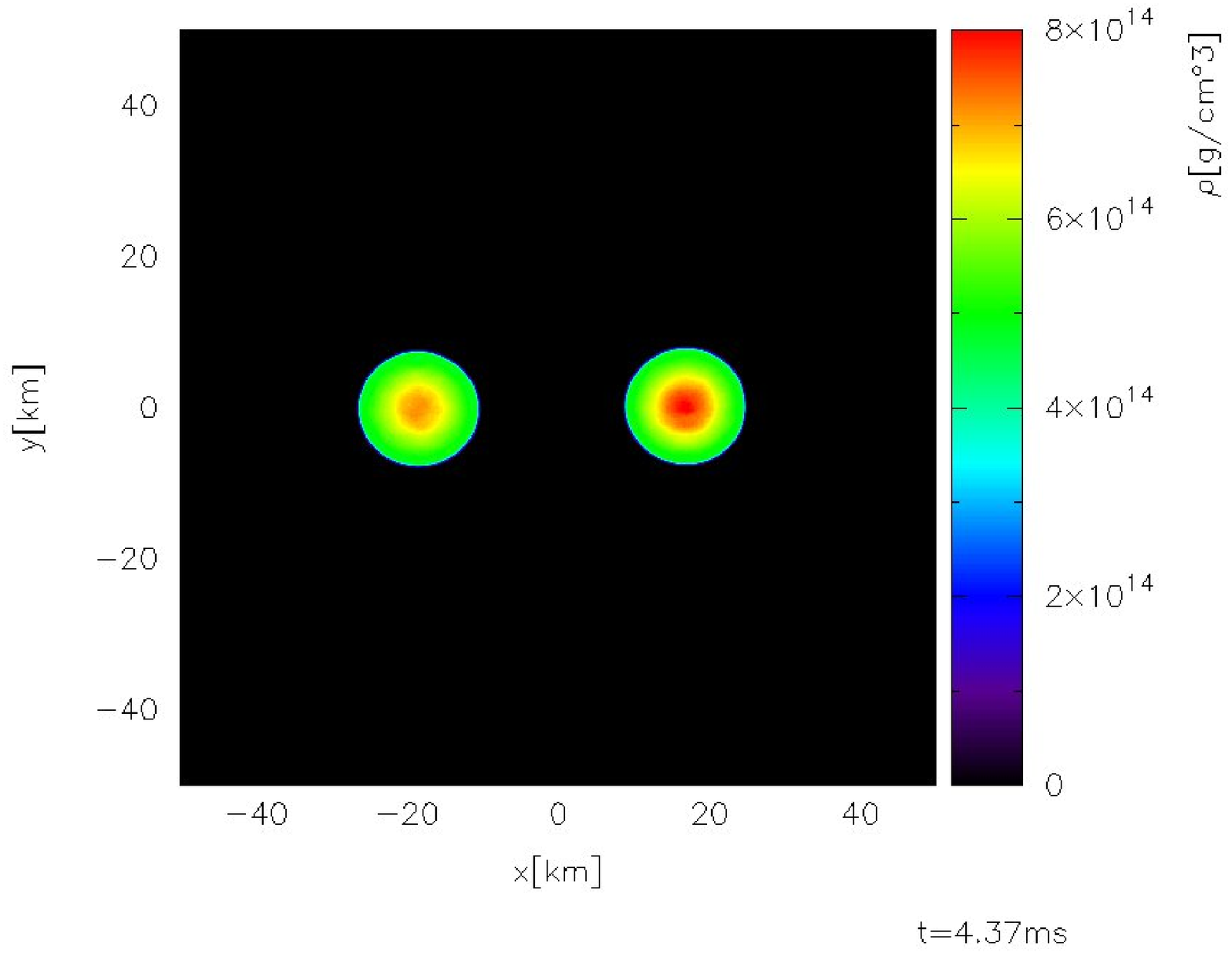}
\epsfxsize=3.4in
\leavevmode
\epsffile{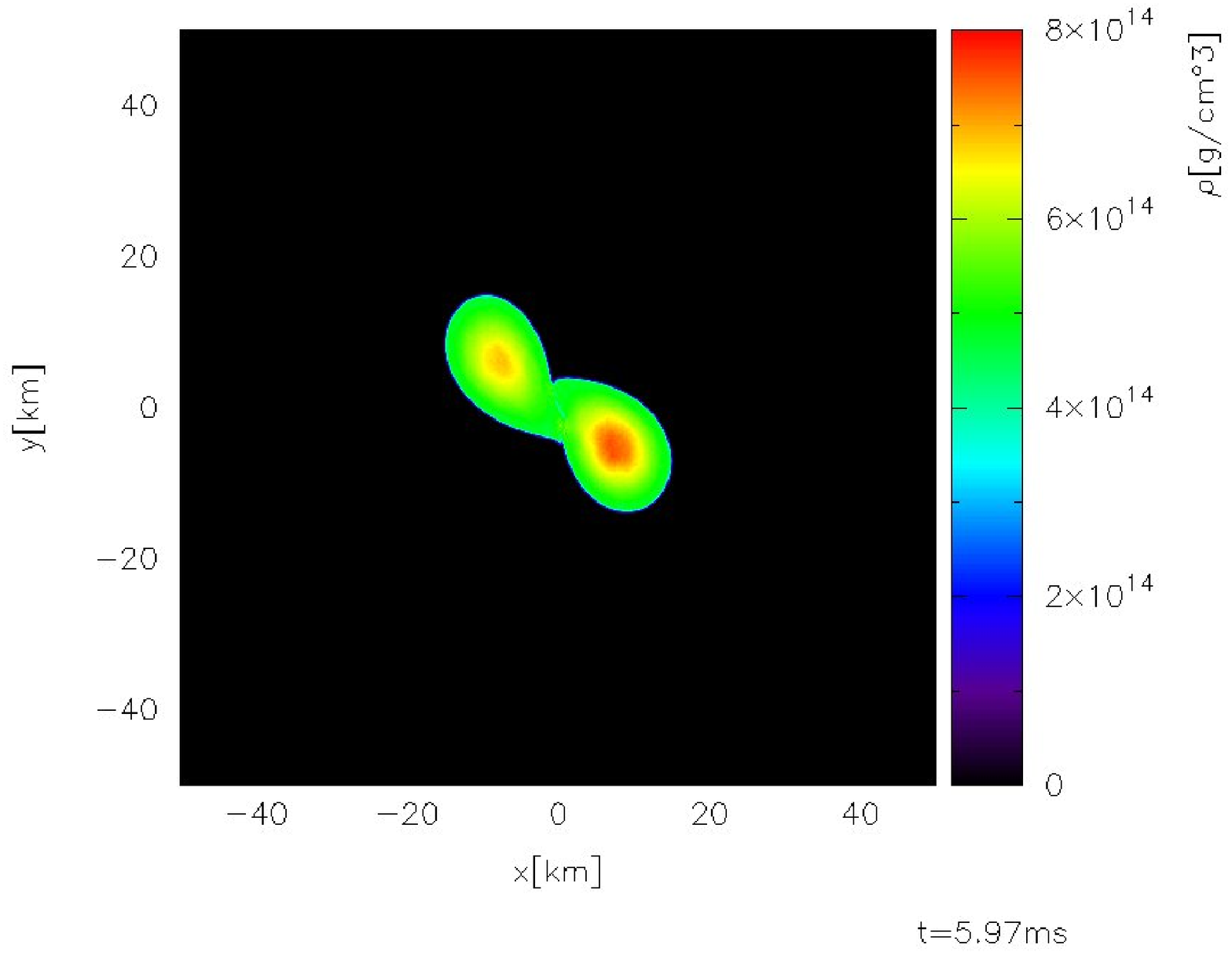}\\
\epsfxsize=3.4in
\leavevmode
\epsffile{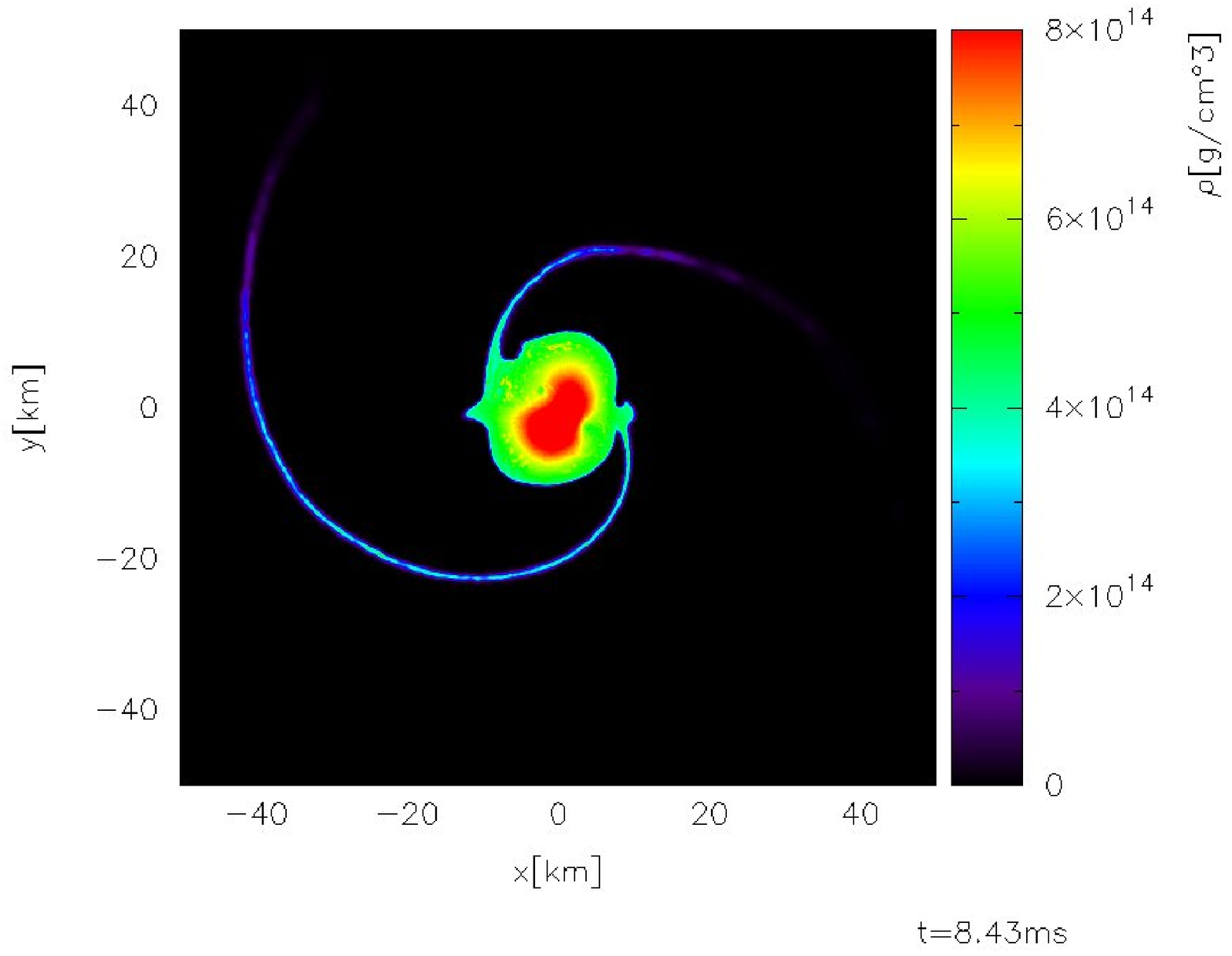}
\epsfxsize=3.4in
\leavevmode
\epsffile{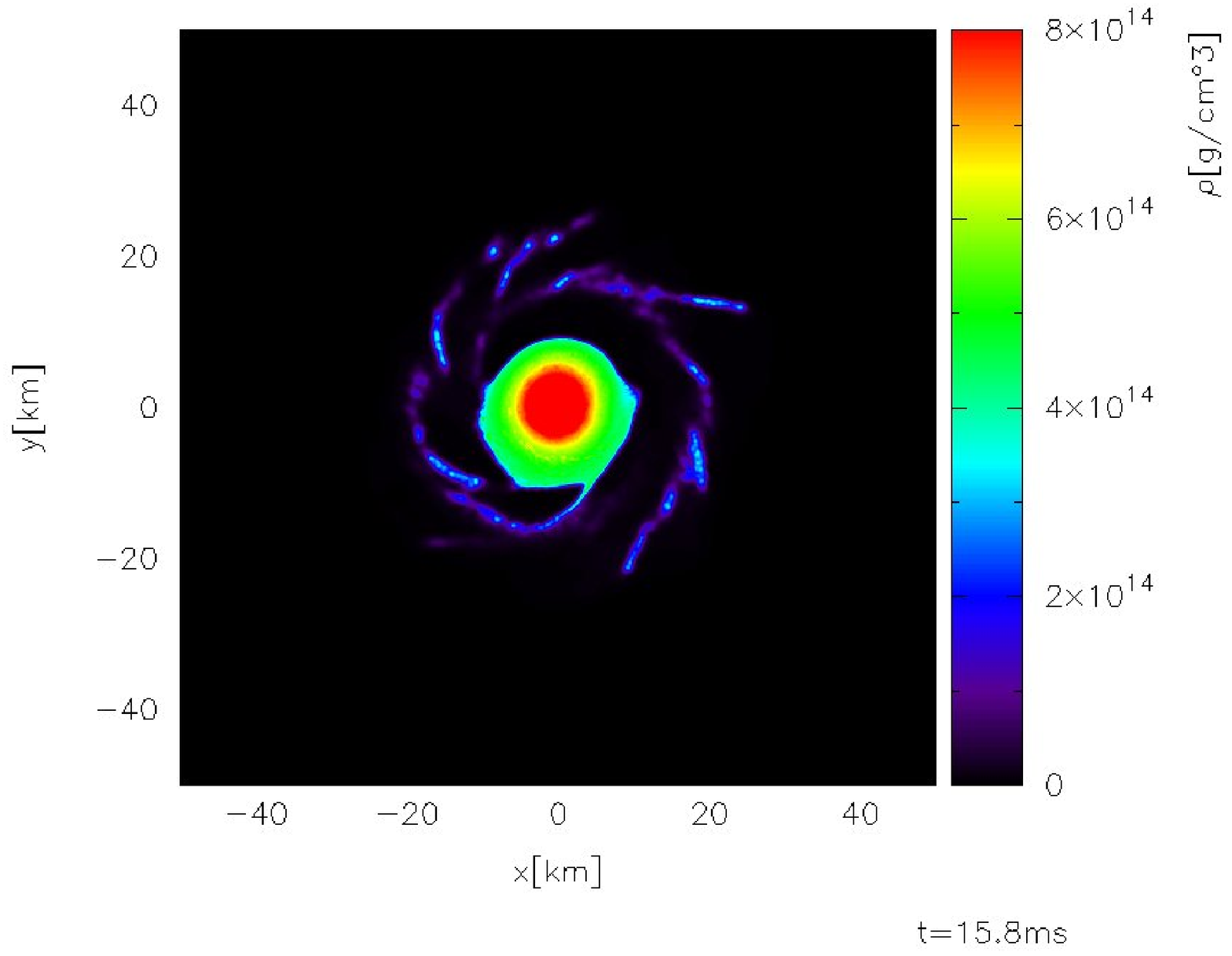}
\vspace{-4mm}
\caption{
\label{fig:snap}Evolution of the rest mass density in the orbital plane of a merging SS binary with 1.2~$M_{\odot}$ and 1.35~$M_{\odot}$ components  for the MIT60. The plots were created with the visualization tool SPLASH \cite{2007PASA...24..159P}.}
\end{figure*}

Figure~\ref{fig:snap} shows the different stages of an SS merger, which finally leads to the formation of such a hypermassive object. The stars have a mass of 1.2~$M_{\odot}$ and 1.35~$M_{\odot}$ and they are described by the MIT60 EoS. The orbital period of the counterclockwise rotating binary shortly before the merger is of the order of 2~ms (upper left panel). Subsequent loss of angular momentum and energy due to GW emission during the inspiral phase leads to a shrinking orbit until the stars finally merge with a relatively big impact parameter (upper right panel). An initially deformed but finally approximately axisymmetric differentially rotating hypermassive object forms, which is stable for more than 10~ms (lower panels) (for a discussion of the delay time $\tau_{\mathrm{delay}}$ to BH collapse see \cite{rezzolla} and Sec.~\ref{ssub:prompt}). After several revolutions of the merger remnant tidal arms form and matter from the tips of these tails becomes gravitationally unbound and so contributes to the cosmic ray flux of strangelets. For the implications of this aspect we refer to \cite{2008arXiv0812.4248B}. Most of the matter in these spiral arms ends up in orbits around the remnant and forms a geometrically thin accretion disk (see Figs.~\ref{fig:snap} and~\ref{fig:snapz}).

One can also see that the matter that is not part of the hypermassive object forms clumps, which is a consequence of the fact that SQM is self-bound. NS mergers instead form dense remnants surrounded by inflated halos with toroidal structures and a smooth density distribution (see Figs.~\ref{fig:snapls} and~\ref{fig:snapz}).

Since we want to discuss in this paper how to discriminate between SS mergers and NS mergers we will concentrate in the following on a selection of models for the LS and MIT60 EoSs and consider general features to reveal fundamental differences. The reason to focus on these two EoSs is that LS as nuclear EoS and MIT60 as SQM EoS are very similar at densities around $10^{15}~\mathrm{g/cm^3}$ and lead to compact stars with similar radii for $M\gtrsim 1.5~M_{\odot}$ (see Figs.~\ref{fig:Prho} and~\ref{fig:MR}). The Shen EoS and the MIT80 EoS exhibit bigger differences and therefore one expects that the distinguishing features are more pronounced. In total, we have computed more then 50 models and discuss in more detail 6 simulations of the 9 listed with their properties in Table~\ref{tab:config}. The choice of three of these binary configurations is motivated by the fact that population synthesis studies \cite{2008ApJ...680L.129B} predict a ``mean'' binary with a total mass of about 2.7~$M_{\odot}$ and a mass ratio close to unity, which is consistent with the masses of the well measured compact star binaries \cite{2004Sci...304..547S}. In addition, we include asymmetric systems with $1.2~M_{\odot}$ and $1.35~M_{\odot}$ to investigate the effects of mass ratios unequal to unity. We choose relatively low masses because only in these cases hypermassive objects can form. Finally, we consider a binary configuration in which the two components have approximately the same radius for the LS and MIT60 EoSs, i.e. we take the stellar masses to be close to the maximum mass for the LS and MIT60 EoSs (see Fig.~\ref{fig:MR}). We will discuss for these models the general dynamics and outcome of the merging phase, the ejecta and torus masses, the influence of thermal effects, and in Sect.~\ref{sec:GW} the features of the GW signals. In particular in the cases where a hypermassive object forms, the GW emission from the postmerger phase yields characteristic information about the EoS of matter at very high densities. 

In Table~\ref{tab:config} we also list results for the Shen EoS without discussing these models in much detail, because the mergers described by this EoS behave qualitatively similar to the ones with the LS EoS. The comparison between different EoSs of NS matter was reported in \cite{2007A&A...467..395O}. We refrain from presenting results for the MIT80 EoS, because all binary configurations considered in Table~\ref{tab:config} form a BH promptly when computed with the MIT80 EoS. However, we will address results with this EoS extensively in Sect.~\ref{sec:GW}.

\begin{table*}
\caption{\label{tab:config} Models that are discussed in the text. $M_1$ and $M_2$ refer to the individual gravitational masses in isolation of the two components of the binary. $f_{\mathrm{max}}$, $f_{\mathrm{peak}}$ and $f_{\mathrm{gap}}$ are characteristic frequencies of the GW signal (see text for definitions). $\tau_{\mathrm{delay}}$ is the delay time between the merging and the formation of a BH. Note that in the cases where the BH collapse does not occur during the simulation, we give a lower limit for $\tau_{\mathrm{delay}}$, this value is determined by the finite simulation time and may differ significantly from the true value (see \cite{rezzolla}). $M_{\mathrm{torus}}$ and $M_{\mathrm{ejecta}}$ are the torus and ejecta masses at the end of the simulations. Note that the torus masses of models LS 1.35$M_{\odot}$+1.35$M_{\odot}$ and Shen 1.35$M_{\odot}$+1.35$M_{\odot}$ are still increasing at the end of the simulations. In the cases we find a vanishing torus mass or ejecta mass we provide the upper bound of this quantity given by the mass resolution of our simulations. $T_{\mathrm{max}}$ denotes the maximal temperature reached during the evolution. For the 1.8$M_{\odot}$+1.8$M_{\odot}$ configurations the presented temperatures are given by the last dump step during the simulations. The actual maximal temperatures during the merging process shortly before the horizon formation are expected to be higher. Table entries in parentheses correspond to models where zero temperature was imposed.}
\begin{ruledtabular}
\begin{tabular}{|l|l|l|l|l|l|l|l|l|l|} 
EoS& $M_1$& $M_2$& $f_{\mathrm{max}}$& $f_{\mathrm{peak}}$& $f_{\mathrm{gap}}$& $\tau_{\mathrm{delay}}$& $M_{\mathrm{torus}}$& $M_{\mathrm{ejecta}}$&$T_{\mathrm{max}}$\\ 
& [$M_{\odot}$]&[$M_{\odot}$]& [kHz]& [kHz]& [kHz]&[ms]&[$M_{\odot}$]&[$M_{\odot}$]&[MeV]\\ \hline
LS& 1.2& 1.35& 1.57 (1.53)& 3.04 (3.16)&2.34& $>$13.1&0.14 (0.19)& 0.008&93\\
MIT60& 1.2& 1.35& 1.80 (1.74)& 3.14 (3.01)&2.59&$>$17.0& 0.08 (0.14)& 0.004&57\\
Shen& 1.2& 1.35& 1.28 & 2.04 & 1.67 & $>$22.0& 0.15 & 0.015&63\\
LS& 1.35& 1.35& 1.75 (1.83)& 3.22 (3.52)&2.76&$>$12.0& 0.04 (0.05)& 0.002&104\\
MIT60& 1.35& 1.35& 1.92 (1.98)& 3.45 (3.37)&2.69&4.37 ($>$16.4)& $\sim$0.09 (0.07)& 0.001&65\\
Shen& 1.35& 1.35& 1.42 & 2.22 & 1.94 & $>$18.1& 0.1 & 0.003&71\\
LS& 1.8& 1.8& 2.16& - &-& 0.42& $<10^{-5}$& 0.0001& 270 \\
MIT60& 1.8& 1.8& 2.20& - &-& 0.48& $<10^{-5}$& $<10^{-5}$& 87 \\
Shen& 1.8& 1.8& 1.62 & - & - & 0.90& 0.004& 0.0002& 221 \\
\end{tabular}
\end{ruledtabular}
\end{table*}

\subsection{Binaries with 1.2~$M_{\odot}$ and 1.35~$M_{\odot}$}
\begin{figure*}[t]
\epsfxsize=3.4in
\leavevmode
\epsffile{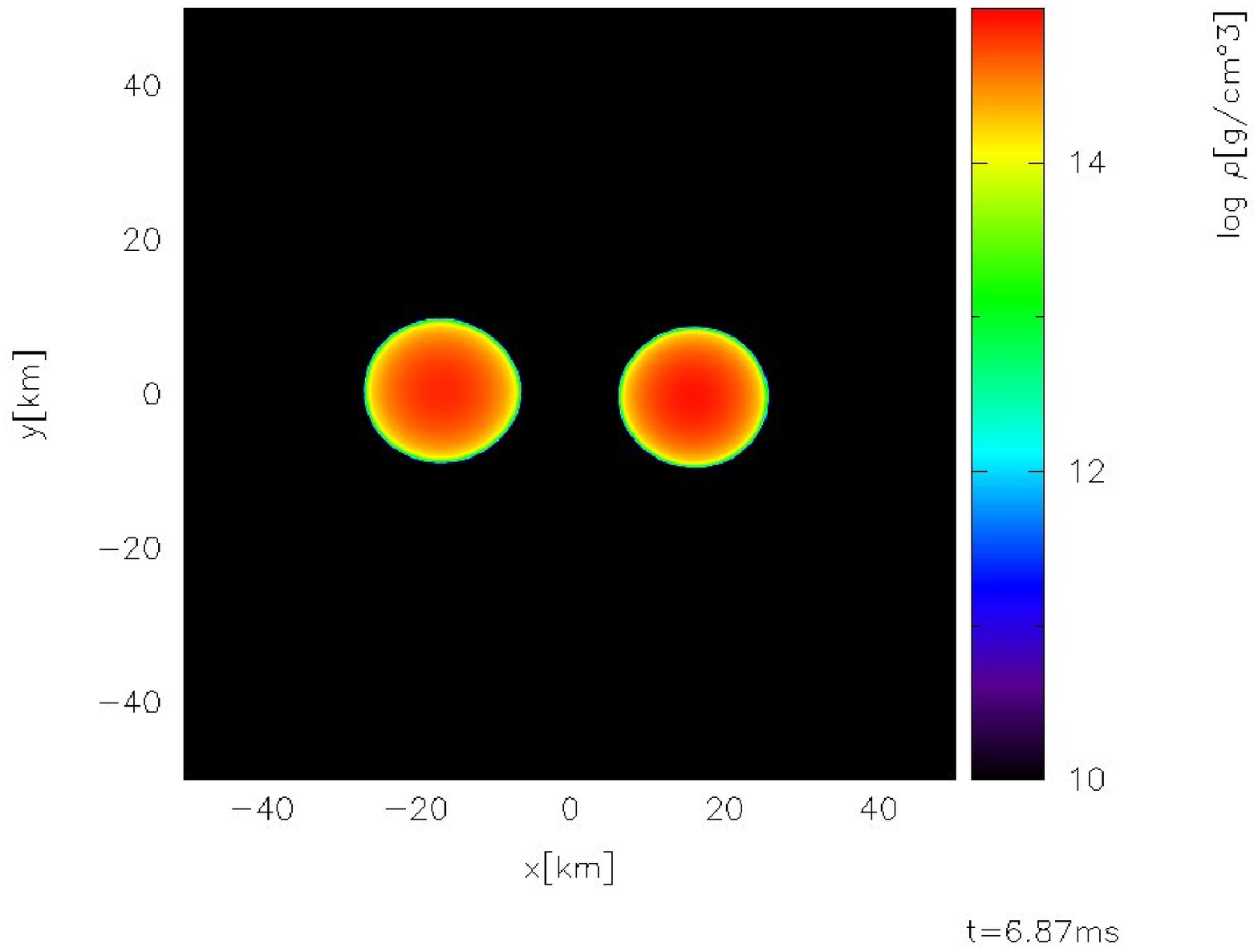}
\epsfxsize=3.4in
\leavevmode
\epsffile{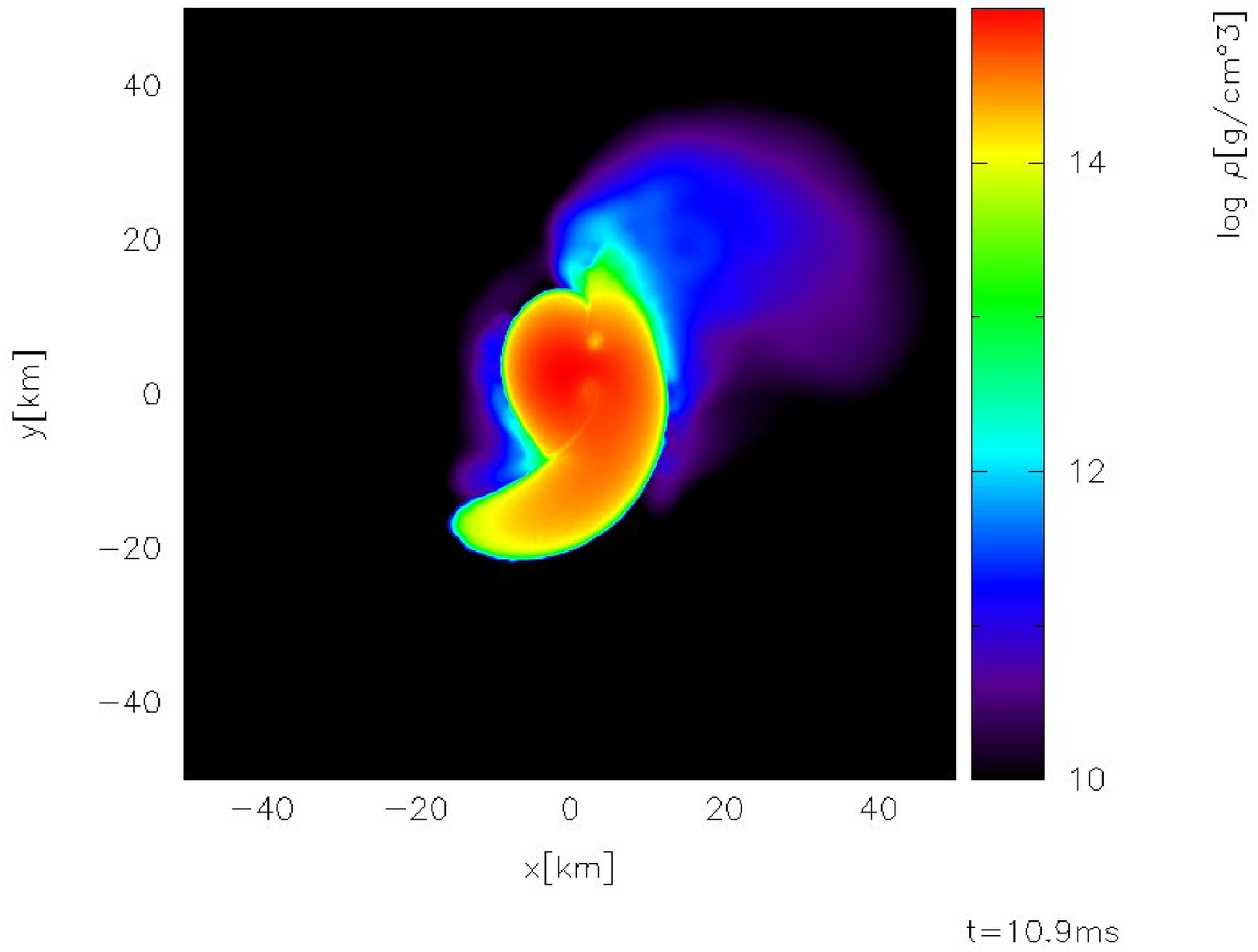}\\
\epsfxsize=3.4in
\leavevmode
\epsffile{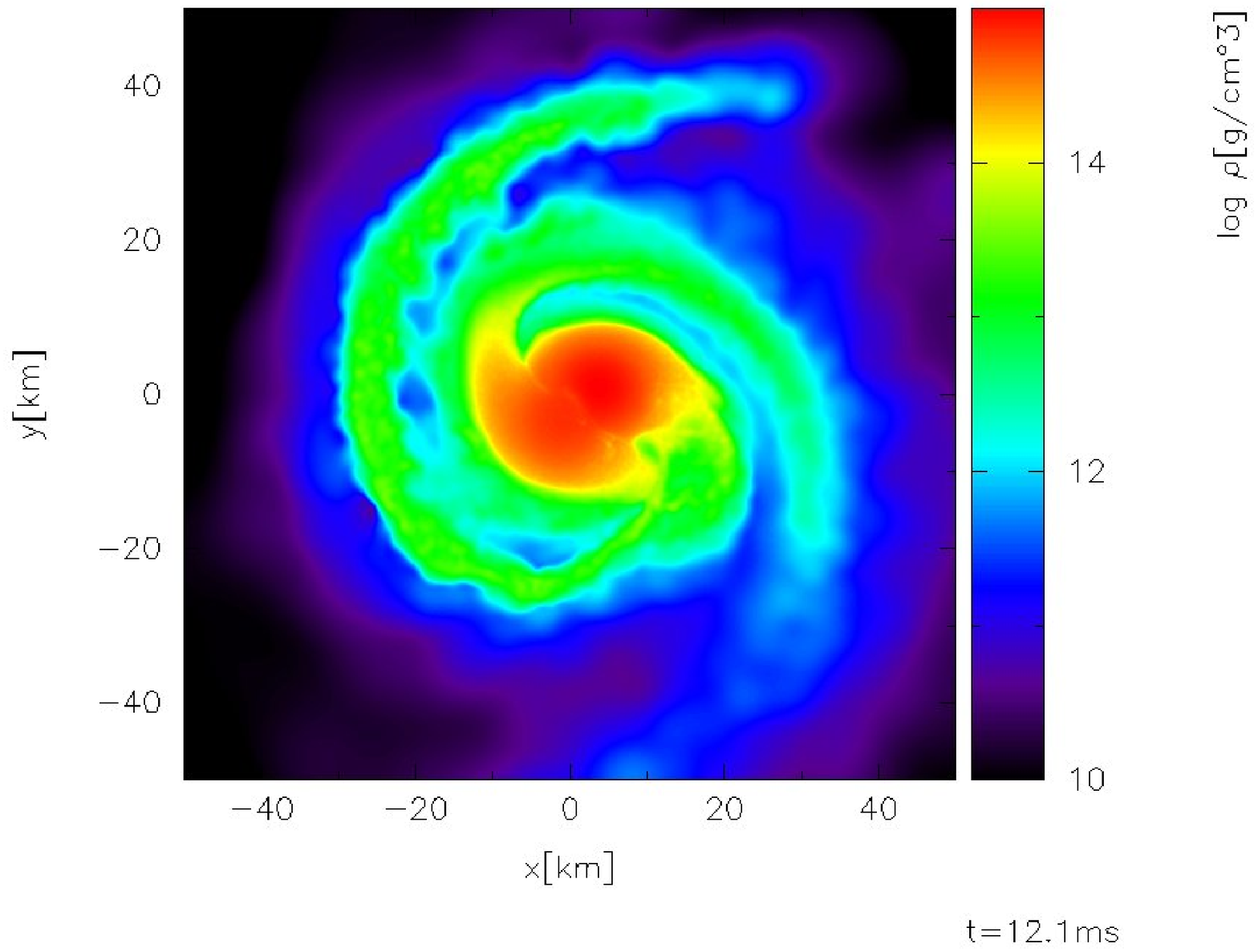}
\epsfxsize=3.4in
\leavevmode
\epsffile{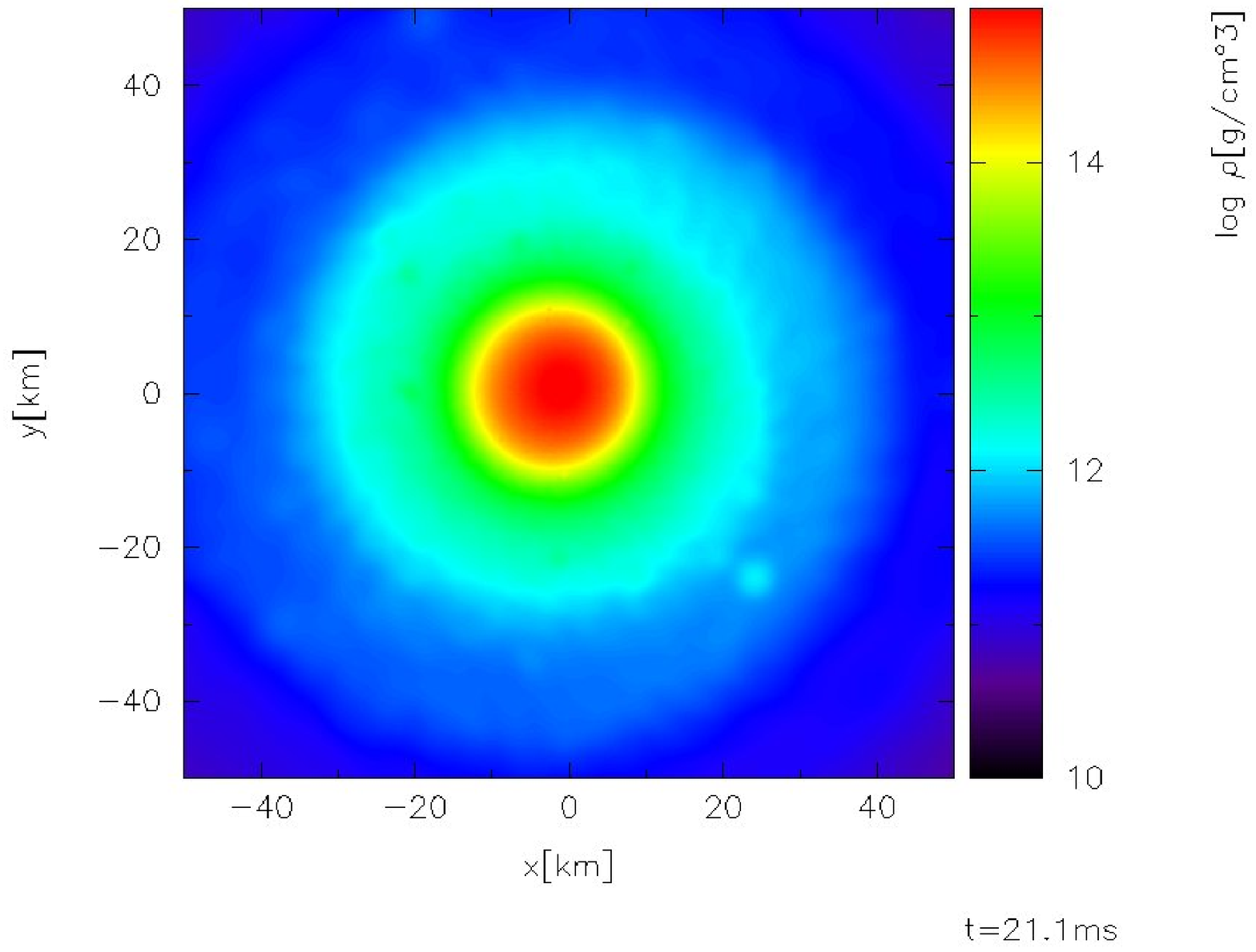}
\vspace{-4mm}
\caption{
\label{fig:snapls}Evolution of the rest mass density in the orbital plane of a merging NS binary with 1.2~$M_{\odot}$ and 1.35~$M_{\odot}$ components for the LS EoS. Note the logarithmic scale of the rest mass density in contrast to Fig.~\ref{fig:snap}. The plots were created with the visualization tool SPLASH \cite{2007PASA...24..159P}.}
\end{figure*}

The two 1.2$M_{\odot}$+1.35$M_{\odot}$ binaries with the LS and MIT60 EoS both form hypermassive objects. The general dynamics of the SS model were described above. Figure~\ref{fig:snapls} shows several stages during the merging of the NS binary to be compared with Fig~\ref{fig:snap}. Different from the SS case, the less massive star is stretched during the final stage of the coalescence and directly forms a massive spiral arm (upper right panel), which means that the lighter star is tidally disrupted during the merging. Continuous mass shedding from this spiral arm feeds an extended toruslike halo around the remnant (lower left panel). The quark stars in contrast, coalesce as entire stars without the prompt formation of such a feature (see Fig.~\ref{fig:disc}).

\begin{figure*}[t]
\epsfxsize=3.4in
\leavevmode
\epsffile{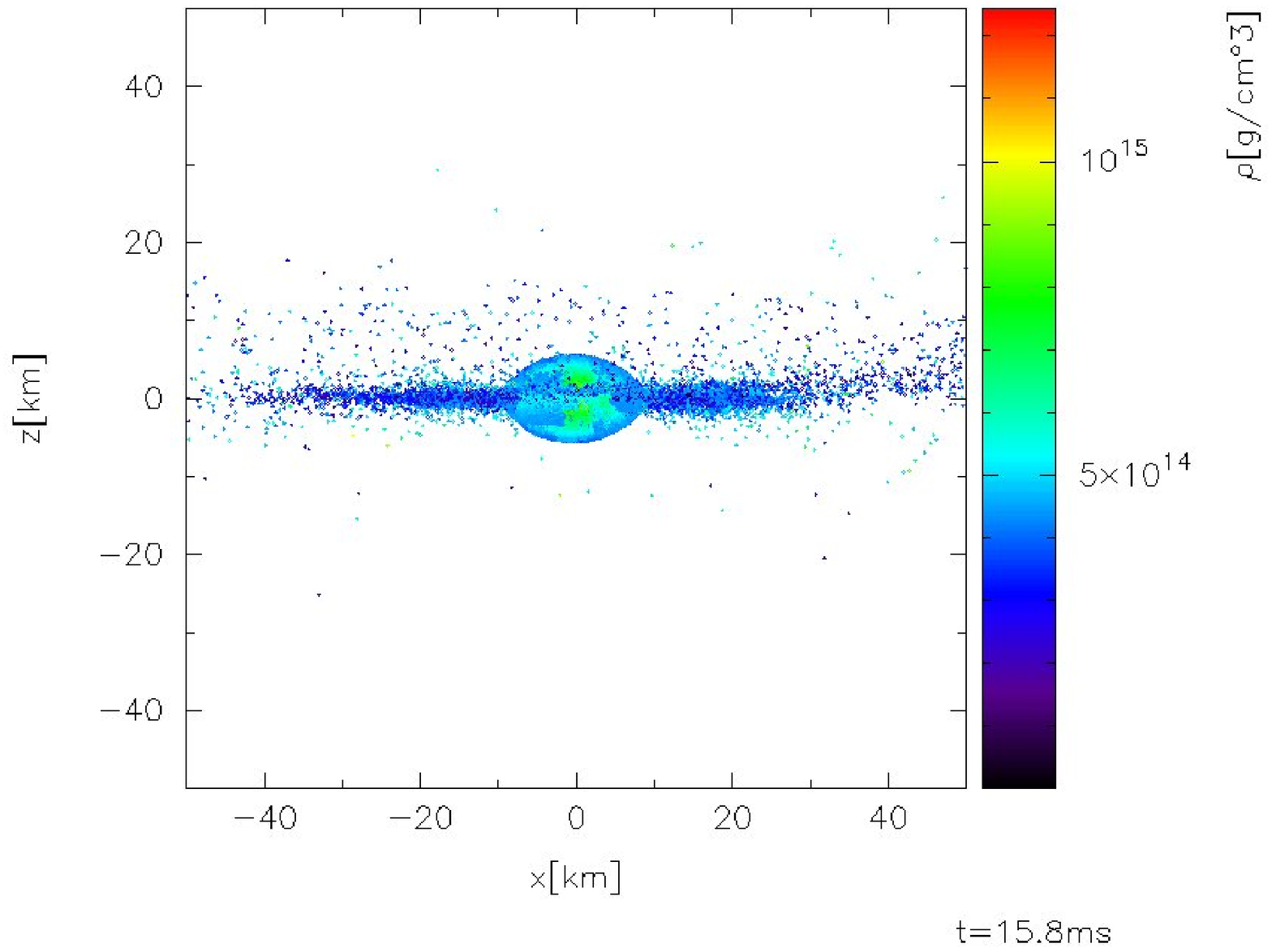}
\epsfxsize=3.4in
\leavevmode
\epsffile{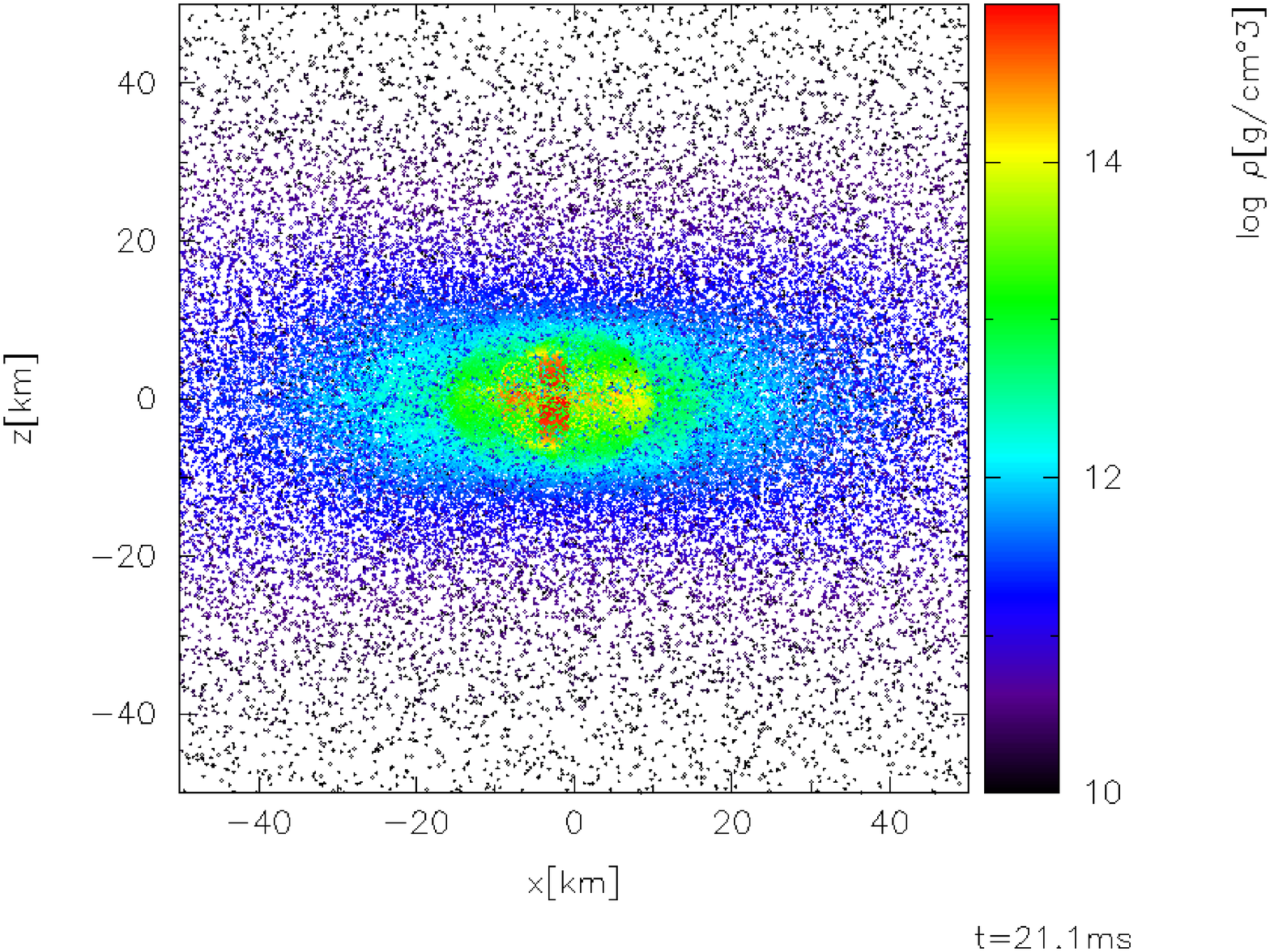}\\
\epsfxsize=3.4in
\vspace{-4mm}
\caption{
\label{fig:snapz} SPH-particle distributions projected onto the x-z-plane perpendicular to the equatorial plane. The left panel shows the SS merger remnant of a 1.2$M_{\odot}$+1.35$M_{\odot}$ binary described by the MIT60 EoS, the right panel the corresponding NS merger remnant described by the LS EoS. Rest mass densities are ascribed to the individual particles and color-coded, but note that due to projection effects foreground particles dominate the visible density structure in the core of the merger remnants. Note the logarithmic scale of the rest mass density for the NS merger remnant. The displayed times correspond to the last panels in Fig.~\ref{fig:snap} and Fig.~\ref{fig:snapls}, respectively. The plots were created with the visualization tool SPLASH \cite{2007PASA...24..159P}.}
\end{figure*}

The structure of the surface layers of the merging stars and of their remnants is fundamentally different. In the case of the LS EoS, matter is also squeezed out from the contact layer, which does not happen for MIT60. Additionally, shortly after the first contact matter streams off from the whole surface of the remnant and fills the diffuse halo (Fig.~\ref{fig:snapz}).

In contrast to the NS case, during the complete merging process of the SSs first the stars then the merged object retain their sharp boundary, no matter is spread into the surroundings. Recall the liquidlike behavior of SQM. Only relatively late after about two revolutions gas is stripped off from the remnant due to the formation of two filigree but very dense spiral arms. This process is constrained to the close vicinity of the equatorial plane (see Fig.~\ref{fig:snapz}) and subsequent fragmentation leads to a disk consisting of clumps of SQM orbiting the compact remnant. Only a small amount of matter from the tips of the tidal tails can escape from the merger site. Remarkably, we find a gap between the merger remnant and the disklike structure (which due to the projection of the particle positions onto the x-z plane is not visible in Fig.~\ref{fig:snapz}). To our knowledge such a distinct feature is unique to SS mergers. The differences of the coalescence and the evolution of the merger remnant can be understood  by means of the higher compactness and self-binding of SQM. We suspect that the reason for the gap is the occurrence of an ISCO outside the surface of the central remnant. An ISCO larger than the stellar radius is known as generic for slowly as well as rapidly uniformly rotating SSs, in particular for supermassive configurations \cite{1999A&A...352L.116S,2000A&A...356..612Z}. Therefore, we expect that this holds also for hypermassive differentially rotating objects.
Note that the omission of a possible nuclear crust of the SSs does not change this picture because the crust would contain only very little mass. A halo fed by this material would be much more dilute than the one in the NS case.

In addition, with the LS EoS the central lapse function decreases during the ringdown of the hypermassive object, while with MIT60 it reaches quickly a nearly constant value as can be seen in Fig.~\ref{fig:rhomax}. This behavior is compatible with the central density, which increases for the remnant consisting of nuclear matter described by the LS EoS (see Fig.~\ref{fig:rhomax}).

Two quantities can be extracted from our simulations for characterizing the mass shedding during the merging: One is the amount of matter that becomes gravitationally unbound, the other is the mass of a torus that will remain after the remnant has collapsed to a BH. The determination of these two quantities is described in \cite{2007A&A...467..395O}. They have direct physical relevance because the ejecta mass is important for nucleosynthesis contributions from NS mergers and for strangelet injection to cosmic rays in the case of SS mergers as discussed in \cite{2008arXiv0812.4248B}. On the other hand, the torus is a potential energy reservoir for powering gamma-ray bursts. The implications of a SQM disk or torus for explaining gamma-ray bursts has not been explored yet. A discussion of this question is beyond the scope of the present paper because it requires an understanding of the neutrino cooling and viscous evolution of the SQM disk.

We find a torus mass of 0.14~$M_{\odot}$ for LS and 0.08~$M_{\odot}$ for MIT60 (see Table~\ref{tab:config}). The differences can be understood by the different dynamical behavior of the merging stars, where tidal disruption favors the formation of a more massive torus. For a detailed discussion of the torus masses and their dependence on the NS merger dynamics we refer the reader to \cite{2006MNRAS.368.1489O}. To clarify this point we show the evolution of the torus mass for both EoSs in Fig.~\ref{fig:disc}. As one can see by the arrows indicating the merger time, the rapid increase for the LS EoS is due to the tidal disruption of the lighter star (compare with Fig.~\ref{fig:snapls}). In contrast, the torus mass of the quark star remnant increases relatively late in the evolution by redistribution of angular momentum to the outer parts of the remnant (compare with Fig.~\ref{fig:snap}).

\begin{figure}
\includegraphics[width=8.9cm]{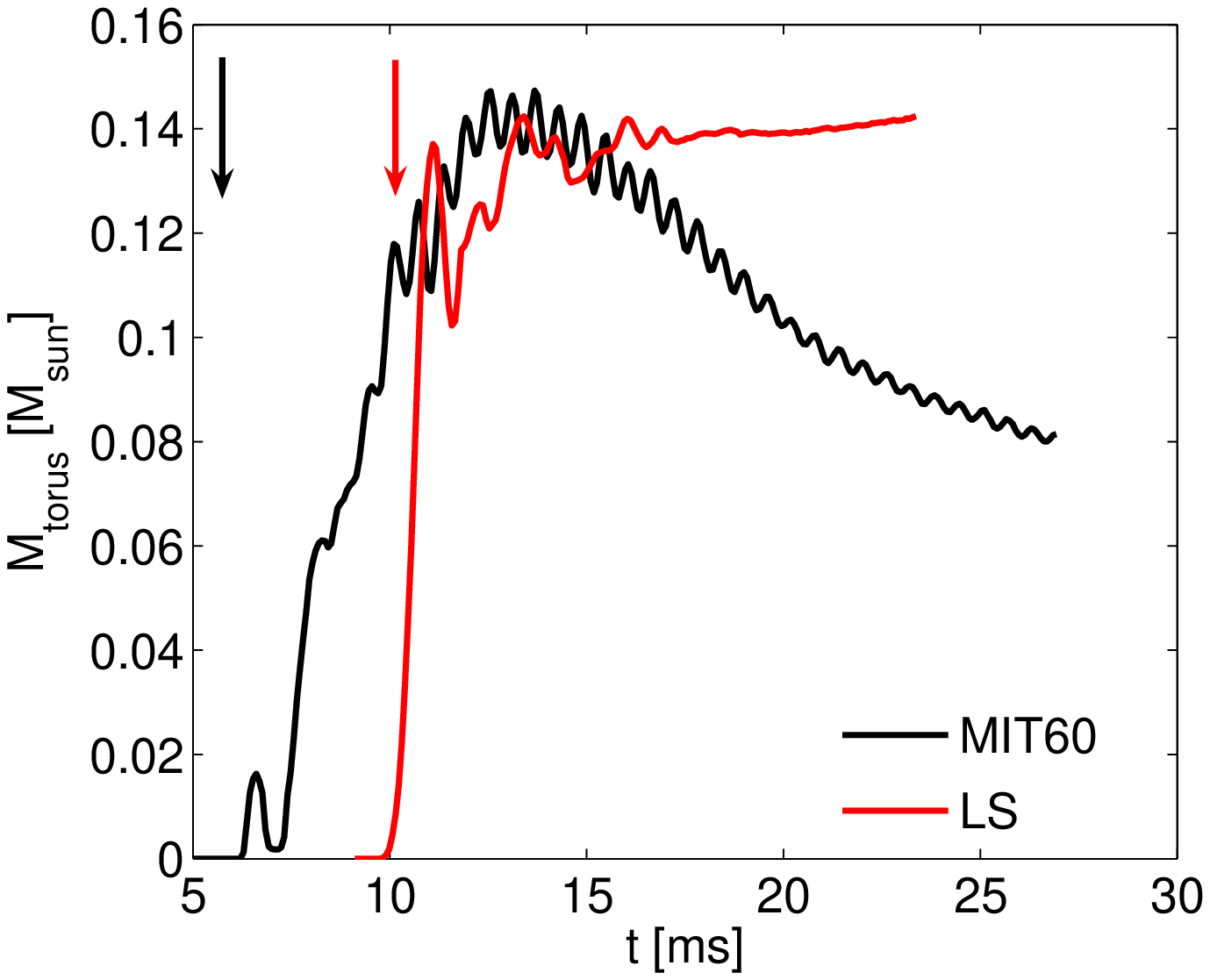}
\caption{\label{fig:disc} Time evolution of the amount of matter fulfilling the torus criterion for the 1.2$M_{\odot}$+1.35$M_{\odot}$ binary mergers with the MIT60 EoS (black) and LS EoS (red). The time labels of Figs.~\ref{fig:snap} and~\ref{fig:snapls} correspond to the time measurement on the horizontal axis. The arrows mark the merging time defined as the moment when the GW amplitude becomes maximal (black for the MIT60 EoS; red for the LS EoS).}
\end{figure}

\begin{figure}
\includegraphics[width=8.9cm]{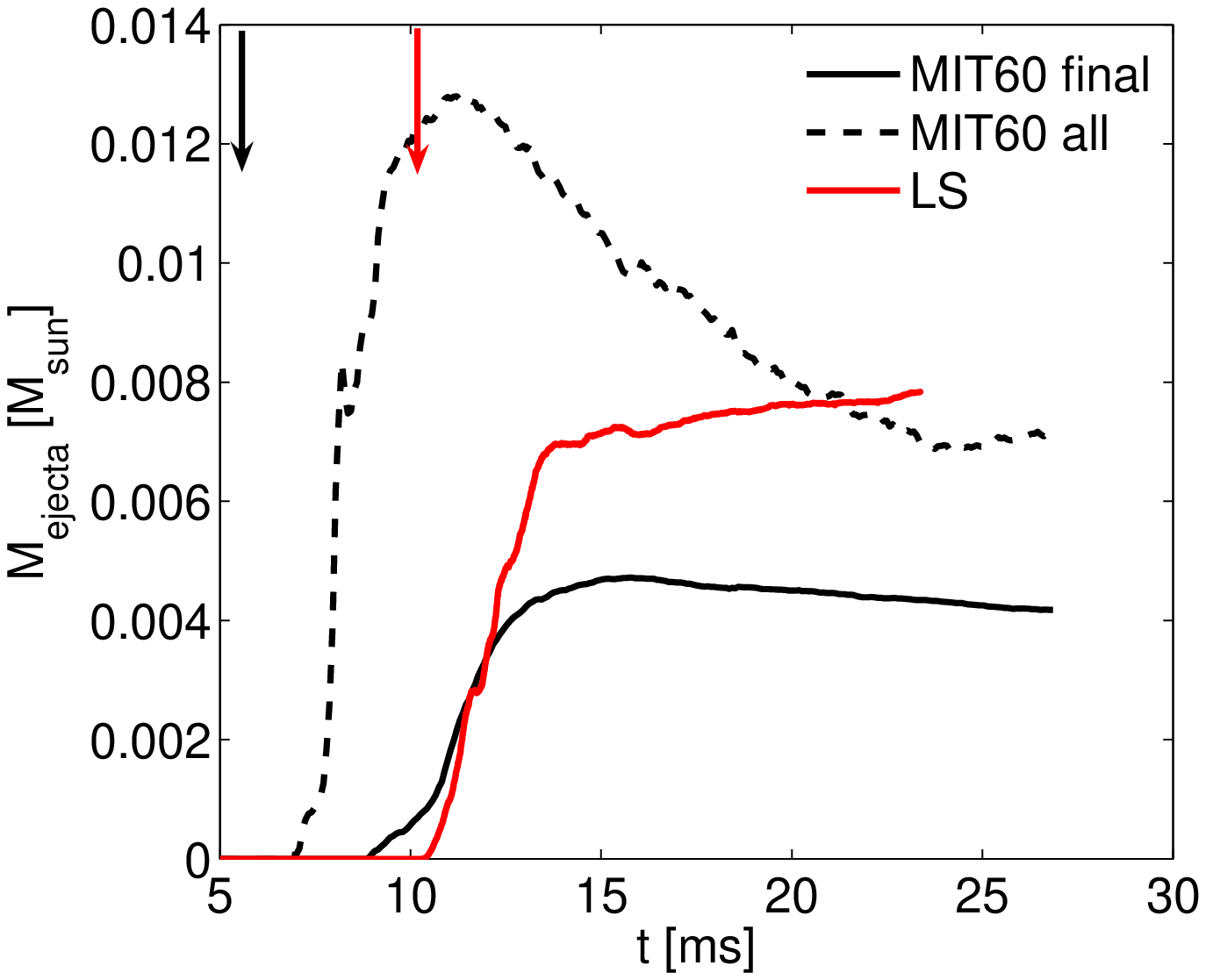}
\caption{\label{fig:ej} Time evolution of the ejecta masses for the 1.2$M_{\odot}$+1.35$M_{\odot}$ binary mergers with the MIT60 EoS (black) 
and LS EoS (red). The time labels of Figs.~\ref{fig:snap} and~\ref{fig:snapls} correspond to the time measurement on the horizontal axis. In the case of the SS merger some SPH particles fulfill the ejecta criterion without being able to leave the merger site. The dashed line gives the total amount of matter fulfilling formally the ejecta condition. The amount of SQM that finally escapes from the merger site is shown by the solid line, where we apply as additional criterion that the SPH particles have a distance of more than 370 km from the remnant. The arrows mark the merging time defined as the moment when the GW amplitude becomes maximal (black for the MIT60 EoS, red for the LS EoS).}
\end{figure}

The ejecta masses are 0.008~$M_{\odot}$ for LS and 0.004~$M_{\odot}$ for MIT60 (Table~\ref{tab:config}). The smaller values in the latter case can be explained as a consequence of the different dynamical behavior due to the higher compactness and the self-binding of the SSs, also impeding the ejection of SQM from the contact layer. This is confirmed by Fig.~\ref{fig:ej} showing the time evolution of the ejecta masses. While during the NS merging matter becomes unbound shortly after the first contact (compare with Fig.~\ref{fig:snapls}), in the SS case the ejecta mass starts growing only about 2~ms after the plunge when the spiral arms form (Figs.~\ref{fig:snap} and~\ref{fig:disc}). Note that not all SPH particles that formally fulfill the ejecta criterion (dashed black curve in Fig.~\ref{fig:ej}) are finally able to escape to infinity because they interact with other particles. Here again the ``stickiness'' of SQM as self-bound fluid manifests itself. By applying an additional distance criterion we estimate the true amount of ejecta (solid black curve in Fig.~\ref{fig:ej})

\subsection{Binaries with 1.35~$M_{\odot}$ and 1.35~$M_{\odot}$}
The merging of two stars with 1.35~$M_{\odot}$ and 1.35~$M_{\odot}$ proceeds differently from the asymmetric case for the nuclear LS EoS. In contrast, in the case of quark-star binaries the merging of symmetric and asymmetric systems is similar, i.e. the stars collide as whole objects with only slight deformation (no tidal disruption). Since the stars possess the same mass, the coalescence and remnant dynamics are symmetric with respect to the two stars. A double-core structure forms and the cores oscillate against each other until the remnant settles to an axisymmetric hypermassive object. However, after about 5 revolutions and several bounces of the two high-density cores, the SS merger remnant collapses to a BH. The time $\tau_{\mathrm{delay}}$ between the plunge (time when the GW amplitude becomes maximal) and the collapse is 4.37~ms. The formation of spiral arms begins after about 2.5 revolutions and ends shortly before the collapse.

Again the way matter is shed off from the merged object is remarkably different. While in the SS case prominent, fine spiral arms develop that are constrained to the vicinity of the equatorial plane, in the NS merger case matter is stripped from the whole surface of the remnant. The spiral arms of the SS merger remnant appear again relatively late and only via these tidal tails matter is carried away from the merger remnant before the latter collapses to a BH. In contrast, the merging NS binary does not develop pronounced spiral arms.

The estimated torus and ejecta masses are given in Table~\ref{tab:config}. The torus mass with the LS EoS is significantly lower (0.04~$M_{\odot}$) than the torus mass with the MIT60 EoS, but it is still increasing at the end of the simulation. The torus mass of the SS merger is not reliable because the collapse to a BH occurs in a phase where stationarity has not been achieved. The value should be considered only as a rough estimate.

The amount of matter that becomes gravitationally unbound from the merger site is 0.002~$M_{\odot}$ for the NS coalescence and 0.001~$M_{\odot}$ in the case of the SS binary. The differences are plausible in view of to the higher compactness of SSs and the special properties of SQM, where additional energy is required to overcome the self-binding. In comparison to the asymmetric binary configurations discussed above, the torus masses as well as the ejecta masses are lower for symmetric mergers.

\subsection{Binaries with 1.8~$M_{\odot}$ and 1.8~$M_{\odot}$}

The last binary configuration that we include in this detailed discussion consists of two stars with the same gravitational mass of 1.8~$M_{\odot}$ and the same radius for both EoSs (about 10.5 km). Both merging binaries promptly collapse into a BH after the stars have come into contact (see Fig.~\ref{fig:rhomax} for the model with the MIT60 EoS). The delay times for that to happen are of the order of half a millisecond (Table~\ref{tab:config}). The stars approach each other symmetrically and they become only slightly deformed before they touch. Note the central decompression phase as reaction to the tidal fields during the merging shortly before the central density increases rapidly (shown for the MIT60 EoS in Fig.~\ref{fig:rhomax}). Also if we cannot follow the complete collapse to a BH, we are able to give an estimate of the torus masses (see \cite{2007A&A...467..395O}). For both EoSs we find a vanishing torus mass. This agreement is reasonable because the mass and radius of the compact objects with both EoSs are the same and the stars are only slightly deformed prior to the collapse. With the LS EoS we find about 0.0001~$M_{\odot}$ that fulfill the criterion for gravitationally unbound matter, but we stress that this number is very uncertain. These ejecta originate from the contact layer between the two stars. In contrast, we do not find any ejecta for MIT60 as during a SS coalescence no matter is squeezed out from the region between the stars. Strange matter becomes only unbound from the tips of tidal tails that form only after several revolutions of the merger remnant. In the case of a prompt collapse, however, this origin is excluded.

\subsection{Thermal effects}
\begin{figure*}
\epsfxsize=3.4in
\leavevmode
\epsffile{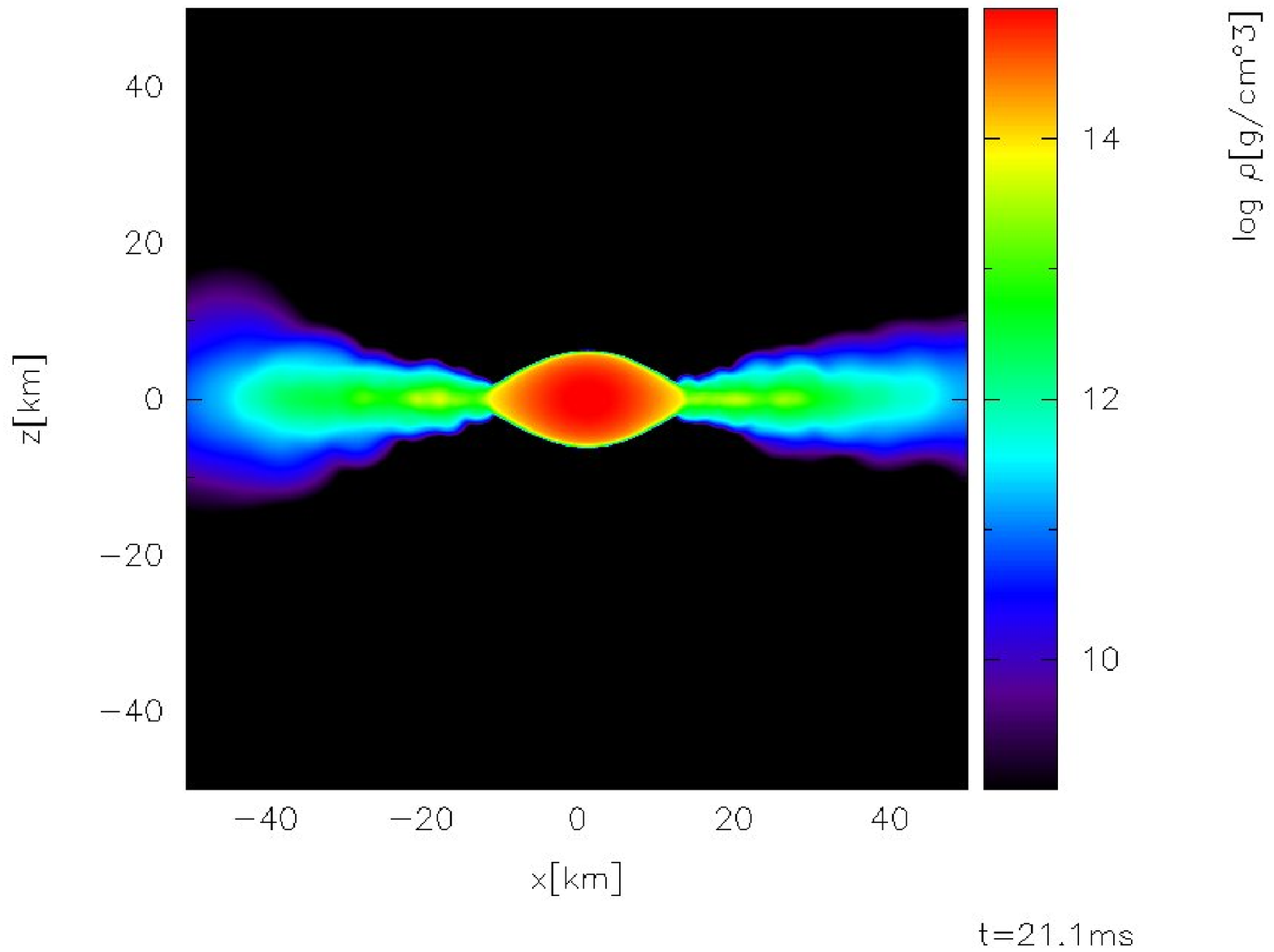}
\epsfxsize=3.4in
\leavevmode
\epsffile{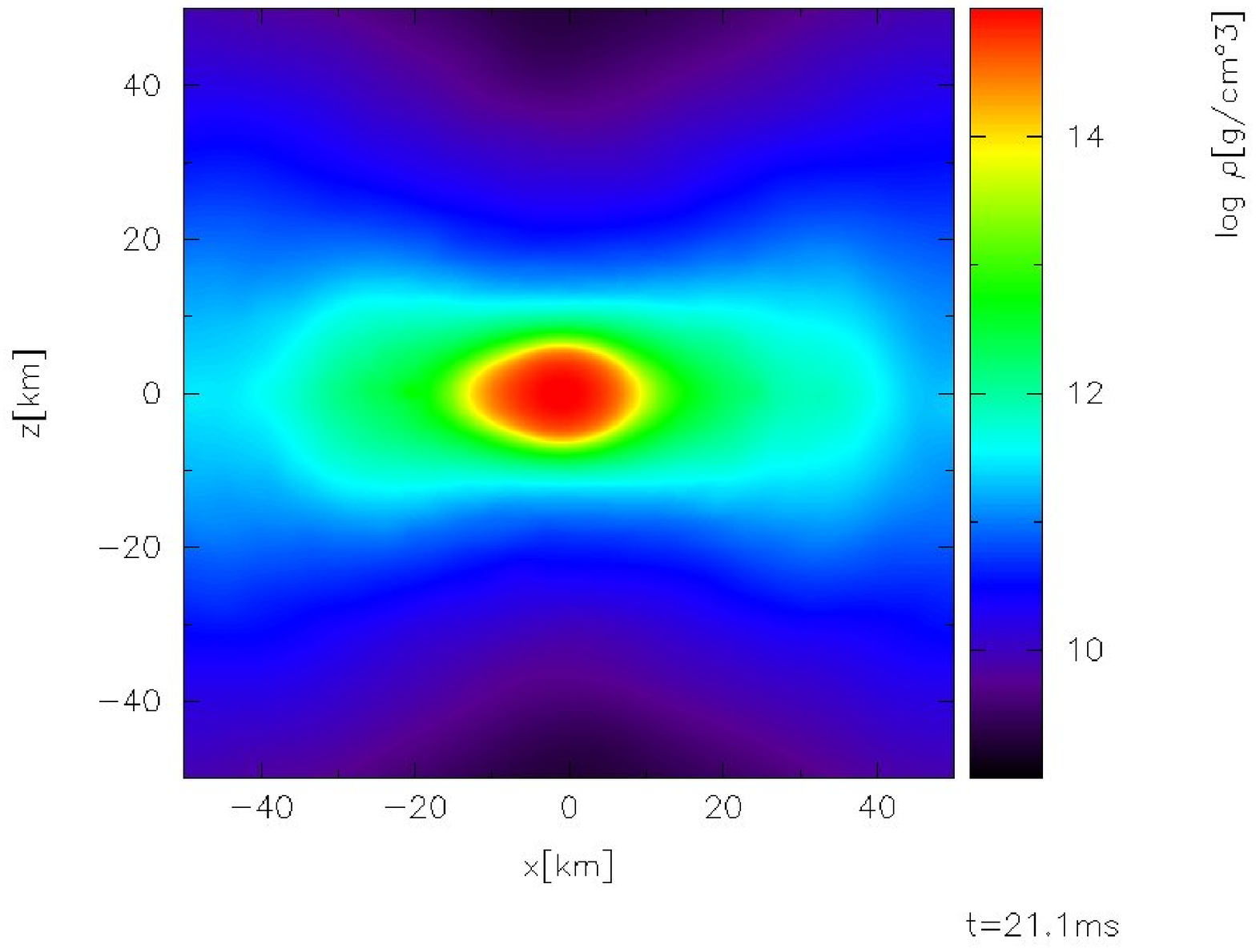}\\
\epsfxsize=3.4in
\vspace{-4mm}
\caption{
\label{fig:snapzthermal} Rest mass density perpendicular to the orbital plane of a merging NS binary with 1.2~$M_{\odot}$ and 1.35~$M_{\odot}$ components for the LS EoS. The left panel shows the results of a simulation where a zero temperature constraint was imposed. The right panel gives the rest mass density of a simulation including thermal effects and corresponds to the lower right panel in Fig.~\ref{fig:snapls} and the right panel in Fig.~\ref{fig:snapz}. The plots were created with the visualization tool SPLASH \cite{2007PASA...24..159P}.}
\end{figure*}

It is known that thermal effects play a role during the merging and postmerging evolution of binary NSs \cite{2007A&A...467..395O}. To which extent this is true in the case of SSs is \it{a priori} \rm not clear because of the different properties of SQM in comparison to NS matter (see Fig.~\ref{fig:Prho}). Temperature effects clearly become important in the contact layer of the colliding stars, where compression is extreme and shocks occur. Because of the differential rotation the hot sheared matter is spread over the entire star forming an inhomogeneous temperature distribution in the remnant. For the MIT60 EoS the maximum temperature reached during the merging of the binary with 1.2~$M_{\odot}$ and 1.35~$M_{\odot}$ stars is about 57 MeV, while the symmetric binary with two stars of 1.35 reaches a maximum temperature of about 65 MeV. In comparison, NS mergers described by the LS EoS yield maximum temperatures of about 93 MeV and 104 MeV for the same binary configurations. For both kinds of matter the highest temperatures are obtained shortly after the first contact of the stars, when the maximal densities peak for the first time (see Fig.~\ref{fig:rhomax}). The higher values in the models with the LS EoS can be understood by the shock heating and the compression of low density material from the surfaces of the initial stars. The surface layers of SSs have already initially supernuclear densities and therefore their relative compression is much smaller.

In order to check the influence of nonzero temperatures we performed simulations setting $T=0$ for the 1.2$M_{\odot}$+1.35$M_{\odot}$ binaries and the 1.35$M_{\odot}$+1.35$M_{\odot}$ binaries. This choice corresponds to the extreme and unrealistic case of a perfectly efficient cooling that carries away all thermal energy instantaneously. A similar approach was chosen in \cite{2008PhRvD..78h4033B} with a polytropic EoS, because effectively, the use of $P=K\rho^{\Gamma}$ with $K, \Gamma=\mathrm{const}$ implies the same physical assumption (``isentropic case''), which was compared to results obtained with an EoS of the type $P=\rho\epsilon$ including thermal effects (``ideal-fluid case''). In Ref.~\cite{2008PhRvD..78h4033B} it was reported that due to the larger pressure support in the ideal-fluid case nonzero temperature leads to a later BH formation than the isentropic treatment. Moreover they found remarkable differences of the remaining tori after the BH collapse had occurred. In particular, in the cases of neglected thermal effects the tori were flat, disklike structures, whereas the mergers described by an ideal-fluid EoS produced inflated tori. For NS models these findings are confirmed by our simulations as displayed in Fig.~\ref{fig:snapzthermal} showing the rest mass density in the x-z-plane perpendicular to the orbital plane for the 1.2$M_{\odot}$+1.35$M_{\odot}$ merger remnant with the LS EoS. In the left panel $T=0$ was imposed, leading to a much smaller vertical extension of the outer remnant parts. Note also that the density in the disk in the equatorial plane is higher by one order of magnitude compared to the density in the ``hot'' torus (right panel). This leads to the conclusion that the additional pressure support by thermal effects is essential for the structure of the remnant. We find about 30\% higher torus masses in our $T=0$ simulation with the LS EoS. We attribute this finding at least partially to the decrease of the gravitational mass due to ``cooling losses'', which is about 0.035~$M_{\odot}$ for this configuration. Lowering artificially the gravitational mass in our torus criterion by this amount of matter increases the estimate for the torus mass in our model with thermal effects. Qualitatively the same behavior is found for the 1.35$M_{\odot}$+1.35$M_{\odot}$ binary with an increase of the torus mass of about 30\% in the $T=0$ simulation compared to the $T\neq 0$ case.

Another interesting effect of the perfectly efficient cooling can be seen in the simulation of the binary with two 1.35~$M_{\odot}$ SSs for the MIT60 EoS. As described above, the merger remnant of this configuration collapses to a BH after $\tau_{\mathrm{delay}}\sim$4~ms. Choosing the same binary setup and the same EoS but imposing the zero temperature constraint during the evolution, the collapse does not occur within the simulation. The high sensitivity of $\tau_{\mathrm{delay}}$ on the total mass of the system (see also \cite{rezzolla}) suggests that this finding can be explained by the lower gravitational mass of the ``cold'' model. The difference in the gravitational mass is about 0.06~$M_{\odot}$ when the remnant of the ``warm'' simulation collapses. Thus, in contrast to NS mergers, where a longer delay timescale for the BH formation was obtained when thermal effects were included \cite{2008PhRvD..78h4033B}, we see the opposite effect in simulations with SQM. This can be understood from the fact that thermal energy adds significantly to the gravitational mass, but hardly affects the pressure and the structure of SSs and their merger remnant (see Fig.~\ref{fig:MR}), because the objects are so dense that typical temperatures do not have a big impact (see Fig.~\ref{fig:Prho}).

These conclusions are confirmed by the structure of the hot merger remnants described by the MIT60 EoS. The hypermassive object of the 1.2$M_{\odot}$+1.35$M_{\odot}$ model (see Fig.~\ref{fig:snap} and left panel of Fig.~\ref{fig:snapz}) looks very similar to the cold one (not shown) and qualitatively similar to the resulting structure of the zero temperature remnants with the nuclear LS EoS (left panel of Fig.~\ref{fig:snapzthermal}). The central cores of the cold and the warm SS merger remnant both have an equatorial diameter of about 20 km and a vertical diameter of about 10 km, which is in accordance with Fig.~\ref{fig:MR}, where warm SQM hardly changes the mass-radius relation of SSs. Both mergers end up with a vertically flat disk structure similar to what is found for models of NS coalescence with thermal effects neglected (left panel of Fig.~\ref{fig:snapzthermal}). As in the NS merger simulations the estimated torus masses are higher for the zero temperature simulations of merging SS (about 100\% for the 1.2$M_{\odot}$+1.35$M_{\odot}$ model; the comparison for the 1.35$M_{\odot}$+1.35$M_{\odot}$ model is not meaningful because of the uncertainties associated with the early collapse of the warm model).

\section{Gravitational waves}
\label{sec:GW}
\subsection{Characteristic features}
\begin{figure}
\begin{center}
\includegraphics[width=8.5cm]{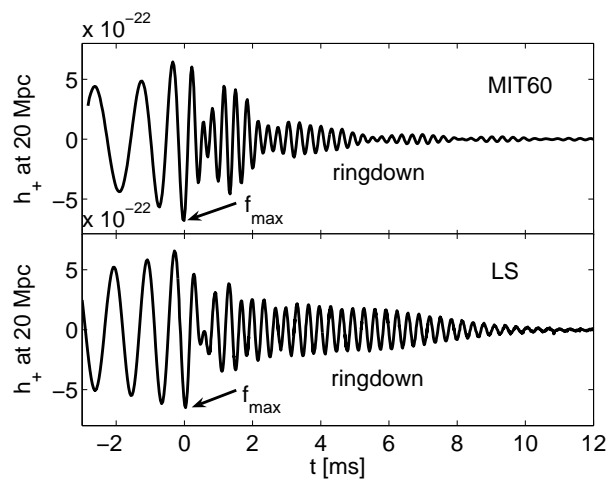}
\caption{Gravitational-wave amplitudes for the plus polarization measured perpendicular to the orbital plane at a distance of 20 Mpc for binary mergers with $M_1=1.2~M_{\odot}$ and $M_2=1.35 ~M_{\odot}$, using the MIT60 EoS (upper panel) and the LS EoS (lower panel).}
\label{fig:gw12135}
\end{center}
\end{figure}

In Fig.~\ref{fig:gw12135} and Fig.~\ref{fig:gw135135} the GW amplitudes of the plus polarization for SS mergers (upper panels) and for NS mergers (lower panels) are shown as measured perpendicular to the orbital plane at a distance $D$ of 20 Mpc. The signals were computed for the models discussed in detail above, thus for mergers of binaries with 1.2~$M_{\odot}$ and~1.35 $M_{\odot}$ stars and for mergers of 1.35$M_{\odot}$+1.35$M_{\odot}$ binaries, both with the MIT60 EoS and the LS EoS. They can be considered as typical of the cases when a hypermassive object forms. When collapse to a BH sets in, our simulation needs to be stopped and the GW signal cannot be tracked to its completion. However, simulations of NS mergers that can follow the formation of the BH reveal that the wave signal is strongly damped in comparison to the hypermassive object case. The only remaining signature after the collapse is associated with the ringdown of the BH, which is fundamentally different, i.e. it has higher frequencies, lower amplitudes and shorter damping times \cite{2006PhRvD..73f4027S,2008PhRvD..78h4033B}.

\begin{figure}
\begin{center}
\includegraphics[width=8.5cm]{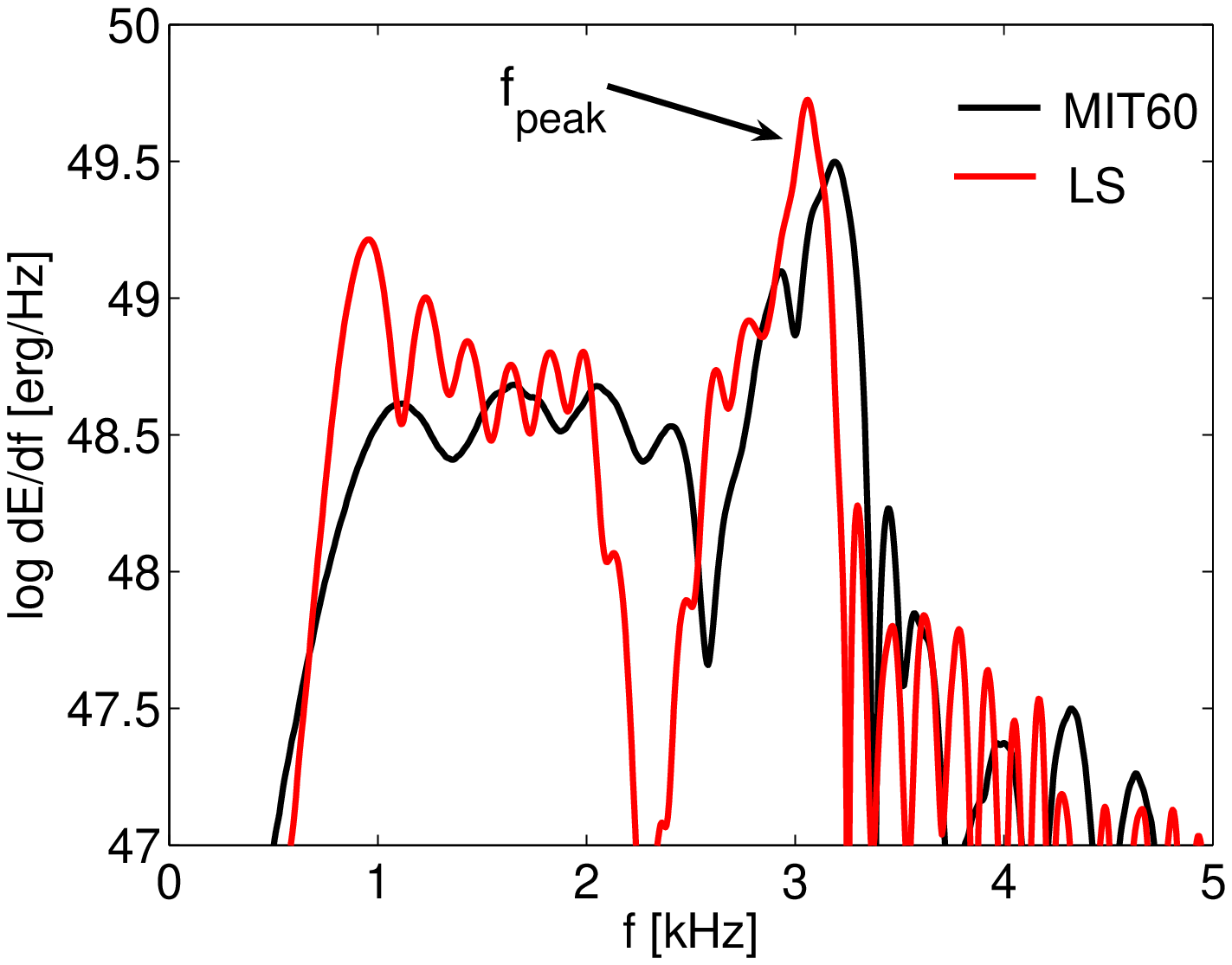}
\caption{Direction and polarization averaged GW luminosity spectra for binaries with $M_1=1.2~M_{\odot}$ and $M_2=1.35~M_{\odot}$ in the case of the MIT60 EoS (black) and the LS EoS (red).}
\label{fig:spec12135}
\end{center}
\end{figure}

\begin{figure}
\begin{center}
\includegraphics[width=8.5cm]{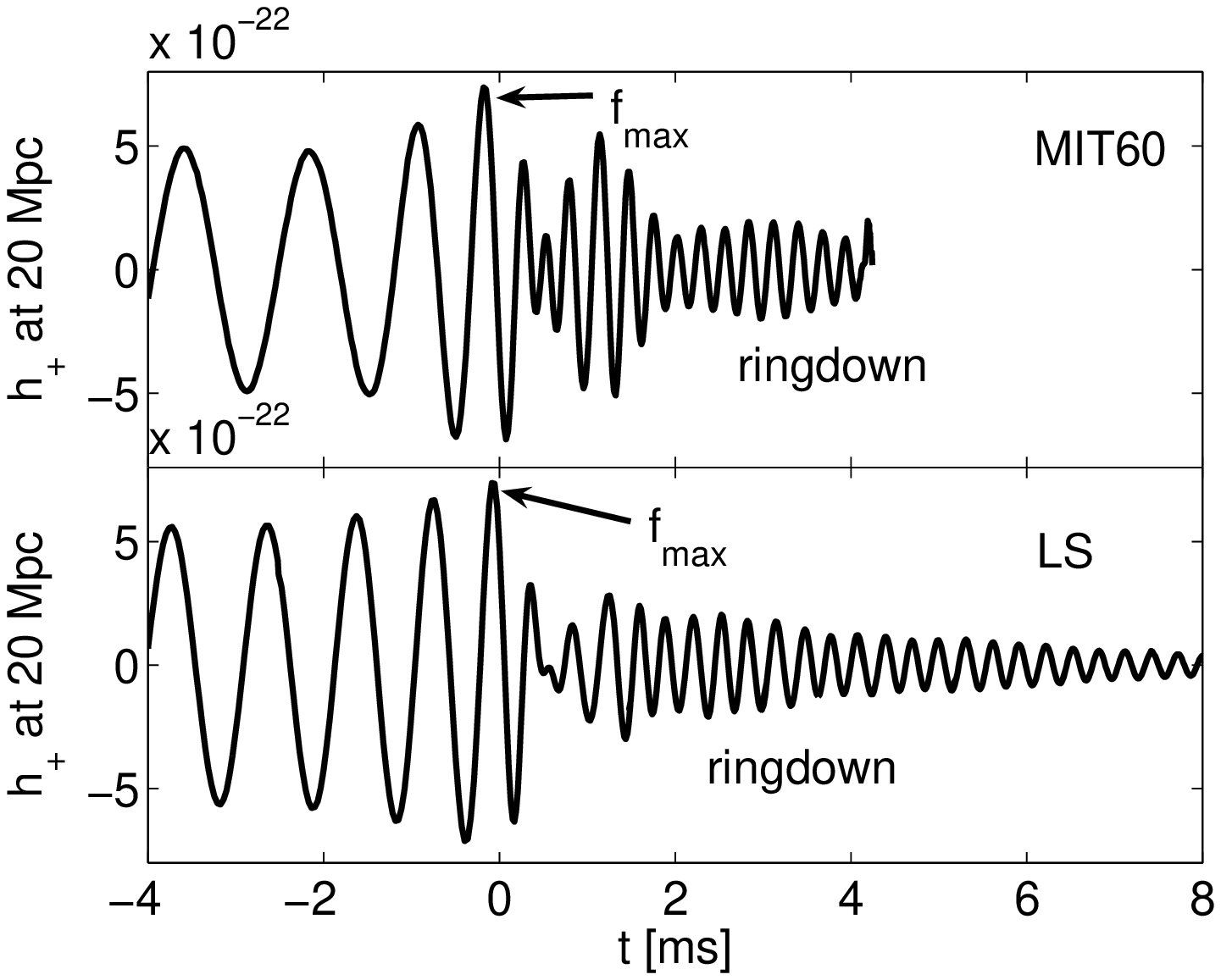}
\caption{Gravitational-wave amplitudes for the plus polarization measured perpendicular to the orbital plane at a distance of 20 Mpc for binary mergers with $M_1=1.35~M_{\odot}$ and $M_2=1.35~ M_{\odot}$, using the MIT60 EoS (upper panel) and the LS EoS (lower panel).}
\label{fig:gw135135}
\end{center}
\end{figure}

\begin{figure}
\begin{center}
\includegraphics[width=8.5cm]{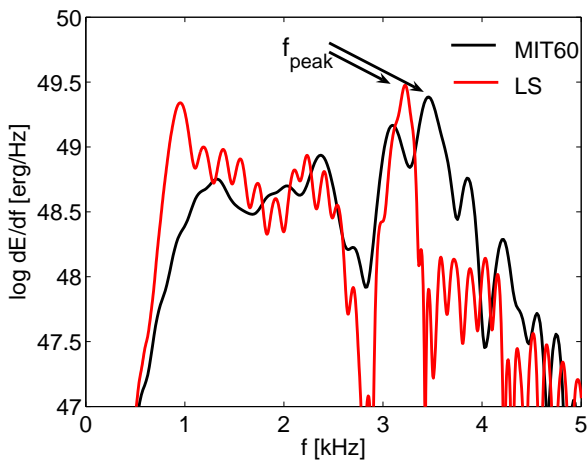}
\caption{Direction and polarization averaged GW luminosity spectra for binaries with $M_1=1.35~M_{\odot}$ and $M_2=1.35~M_{\odot}$ in the case of the MIT60 EoS (black) and the LS EoS (red).}
\label{fig:spec135135}
\end{center}
\end{figure}

The waveforms in Figs.~\ref{fig:gw12135} and~\ref{fig:gw135135} are extracted by an expression of the quadrupole formula that takes into account post-Newtonian effects \cite{1990MNRAS.242..289B}.

The inspiral phase showing an increase of the frequency and of the amplitude can be clearly identified. Its basic signal properties can be easily understood by an analytical model of orbiting point masses. The corresponding wave train of the cross polarization looks similar but is phase-shifted by $\pi/2$. A characteristic quantity that can be extracted from the inspiral phase is the maximum frequency at the end of this stage just before the merging of the two compact stars. More precisely, we define $f_{\mathrm{max}}$ as the frequency at the moment when the amplitude becomes maximal. Note that we shift the time axis to match $t=0$ at this point.

The final plunge produces a rather complicated wave pattern. After it has taken place, one can see a quasiperiodic signal from the ringdown of the postmerger remnant. This oscillation is reflected in the luminosity spectra. In Figs.~\ref{fig:spec12135} and~\ref{fig:spec135135} we show the direction and polarization averaged luminosity spectra $dE_{GW}/df=2\pi ^2 D^2 f^2 \langle|\tilde{h}_+|^2+|\tilde{h}_\times|^2\rangle$ as computed from the Fourier transformed waveforms $\tilde{h}$ for both polarizations. The very pronounced peak (or even multiple peaks) corresponds to the ringdown frequency of the postmerger remnant. This peak is located at a frequency $f_{\mathrm{peak}}$. Together with the maximum frequency during inspiral we now have at hand two characteristic features of the GW emission of compact object mergers. For a discussion and introduction of these quantities in the context of NS mergers, see \cite{2007PhRvL..99l1102O}, where also the prospects of a detection and determination of these characteristics by GW detectors is briefly addressed. Here we will explore to what extent these features can serve to distinguish NS mergers from SS mergers. It should be mentioned that the low-frequency part of our spectra ($\lesssim 1 ~\mathrm{kHz}$) is not reliable because our simulations start only a few orbits before the plunge and therefore power in the low-frequency domain from the preceding inspiral phase is missing. In the Newtonian limit the shape of the GW energy spectrum for the inspiral phase is given by a $f^{-1/3}$ power law.

Comparing the results for the binary with the 1.2~$M_{\odot}$ and 1.35~$M_{\odot}$ components one finds that the maximal frequencies $f_{\mathrm{max}}$ that are reached during the inspiral are 1.80 kHz for MIT60 and 1.57 kHz for LS (Fig.~\ref{fig:gw12135}). The lower value of the NS model appears reasonable considering the lower compactness of the initial stars compared to the quark stars. The ringdown signal after the merging decays more slowly for the nuclear EoS (Fig.~\ref{fig:gw12135}). We attribute this to the higher asymmetry of the merger remnant caused by the tidal disruption and suspect that the persistent oscillations may be supplied by the continued contraction of the remnant (Fig.~\ref{fig:rhomax}). Note also that the amplitude shortly after the merging is  higher for the SS merger and one can clearly recognize a persistent modulation of the signal. The origin of this low-frequency feature is unclear to us and should be examined by a mode analysis of the postmerger remnant.

The spectra look qualitatively similar and in particular the peak frequencies of the postmerger ringdown are very close, both about 3 $\mathrm{kHz}$ (3.04 kHz for the NS binary and 3.14 kHz for the SS binary). A remarkable difference is the pronounced gap below the peak frequency in the spectrum of the case with the LS EoS. Also the reason for this feature should be investigated in detail by a mode analysis. In general, a more detailed analysis of the spectra should reveal some more, less obvious differences. However, this would require a detailed understanding of the mode excitation during the merging, which is beyond the scope of this work. It is also questionable whether such secondary differences can be of observational relevance for detectors available in the near future.

In the case of symmetric configurations with two 1.35~$M_{\odot}$ stars the maximal frequencies during inspiral are 1.92 kHz for MIT60 and 1.75 kHz for the nuclear LS EoS, and as before the GW amplitude shortly after the merging is higher for the MIT60 EoS, while the signal for the LS EoS is weaker but decreases more slowly (Fig.~\ref{fig:gw135135}). A low-frequency modulation of the SS merger signal occurs similar to the waveform found for the asymmetric SS binary. As mentioned above, the merger remnant for MIT60 collapses after some revolutions and so the signal cannot be followed by our simulation any more. Nevertheless, the GW spectrum for the SS merger exhibits more power at high frequencies than the NS merger emission and as in the asymmetric case the peak frequency of the SS merger is slightly higher. Again one recognizes a deep trough in the spectrum of the model with the LS EoS. Although present in all of our models with the LS EoS, it is unclear whether this is a universal feature and if it can be used to distinguish SS mergers from NS mergers in general (for the simulations employing the Shen EoS we also find a pronounced gap feature in the spectra except for two binary configurations; mergers described by the MIT80 EoS only form hypermassive remnants for $M_{\mathrm{tot}}\le2.35~M_{\odot}$).

Since the binaries containing two stars with 1.8~$M_{\odot}$ collapse promptly to BHs for both EoSs, one can only consider the maximal inspiral frequency as characteristic quantity. For the LS EoS one finds $f_{\mathrm{max}}=2.16 ~\mathrm{kHz}$, while for the SS binary the maximal frequency is 2.20~kHz. Since the stars of this binary setup have the same radius for both EoSs, the similarity of these values is not unexpected. In \cite{2009arXiv0901.3258R} it was suggested that the mass-radius relation and thus the EoS of compact stars can be determined from deviations of the GW signal from the point-particle behavior. However, this proposal relies on the close relation between the signal properties and just the stellar radius as the crucial stellar parameter, and neglects the potential influence of a different inner structure of compact stars with the same mass and radius. The same values of $f_{\mathrm{max}}$, one of the most remarkable features of the inspiral phase of our 1.8$M_{\odot}+$1.8$M_{\odot}$ models, confirms that the inner structure has indeed a minor influence even on the dynamics of the late stages of the inspiral phase. Also in our study we find that the mass-radius relation is the EoS property that most sensitively determines $f_{\mathrm{max}}$, because for an NS merger with the stiffer Shen EoS, where a 1.8~$M_{\odot}$ NS has a 5 km bigger radius, we obtain $f_{\mathrm{max}}=1.62 ~\mathrm{kHz}$. On the other hand, a disadvantage of the insensitivity to the inner structure may be that it requires several ``radius measurements'' to constrain the EoS, because one detection may not provide unambiguous information about the EoS. 

\subsection{Binary parameter dependence}

\begin{figure}
\begin{center}
\includegraphics[width=8.5cm]{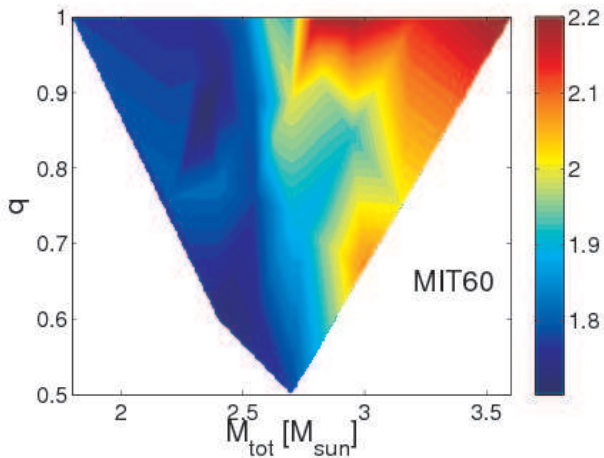}
\caption{Maximal frequency during inspiral color-coded in kHz as a function of the total binary mass $M_{\mathrm{tot}}$ and of the mass ratio $q$ for binary mergers with the MIT60 EoS.}
\label{fig:fmax60}
\end{center}
\end{figure}

\begin{figure}
\begin{center}
\includegraphics[width=8.5cm]{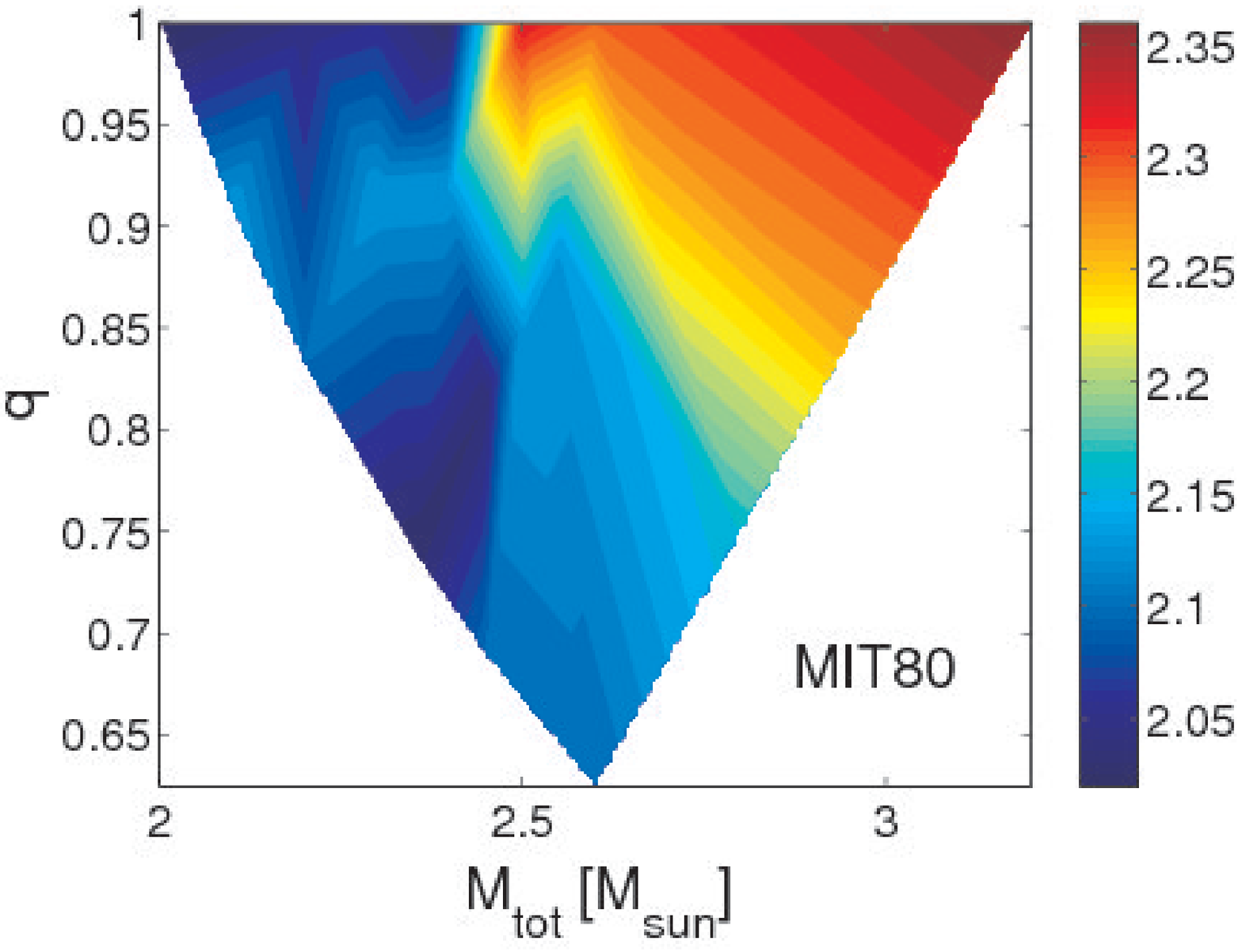}
\caption{Maximal frequency during inspiral color-coded in kHz as a function of the total binary mass $M_{\mathrm{tot}}$ and of the mass ratio $q$ for binary mergers with the MIT80 EoS.}
\label{fig:fmax80}
\end{center}
\end{figure}

\begin{figure}
\begin{center}
\includegraphics[width=8.5cm]{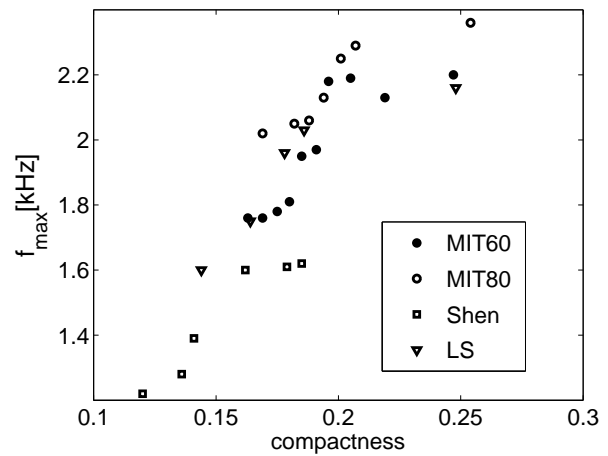}
\caption{ Maximal frequencies during the inspiral phase as a function of the compactness of the initial stars for all configurations with a mass ratio of $q=1$ and all EoSs used in this study. The compactness is defined as the gravitational mass of a single star in isolation divided by its radius in Schwarzschild coordinates.}
\label{fig:fmaxcomp}
\end{center}
\end{figure}

\begin{figure}
\begin{center}
\includegraphics[width=8.5cm]{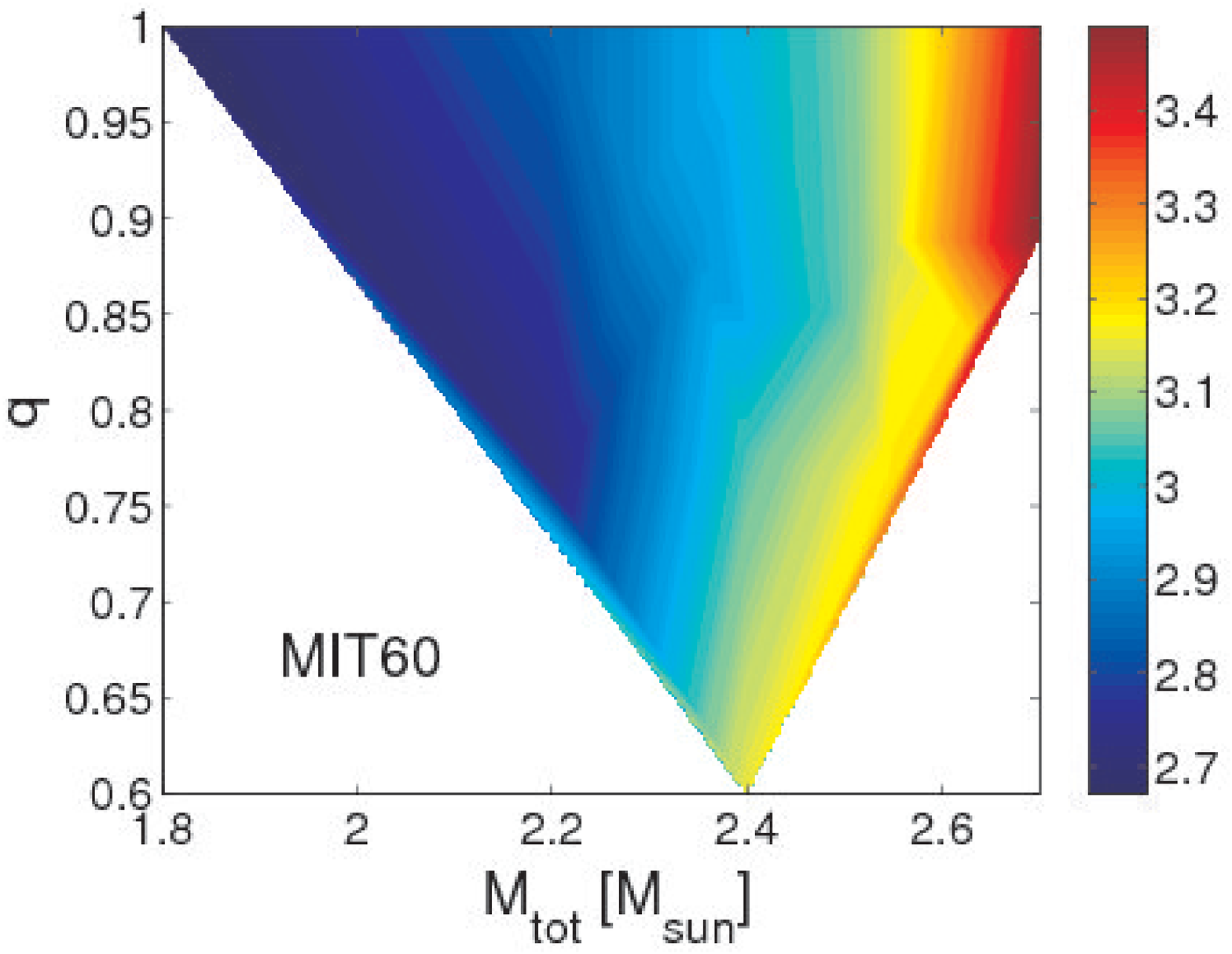}
\caption{Peak frequency of the postmerger ringdown color-coded in kHz as a function of the total binary mass $M_{\mathrm{tot}}$ and of the mass ratio $q$ for binary mergers with the MIT60 EoS.}
\label{fig:fpeak60}
\end{center}
\end{figure}

\begin{figure}
\begin{center}
\includegraphics[width=8.5cm]{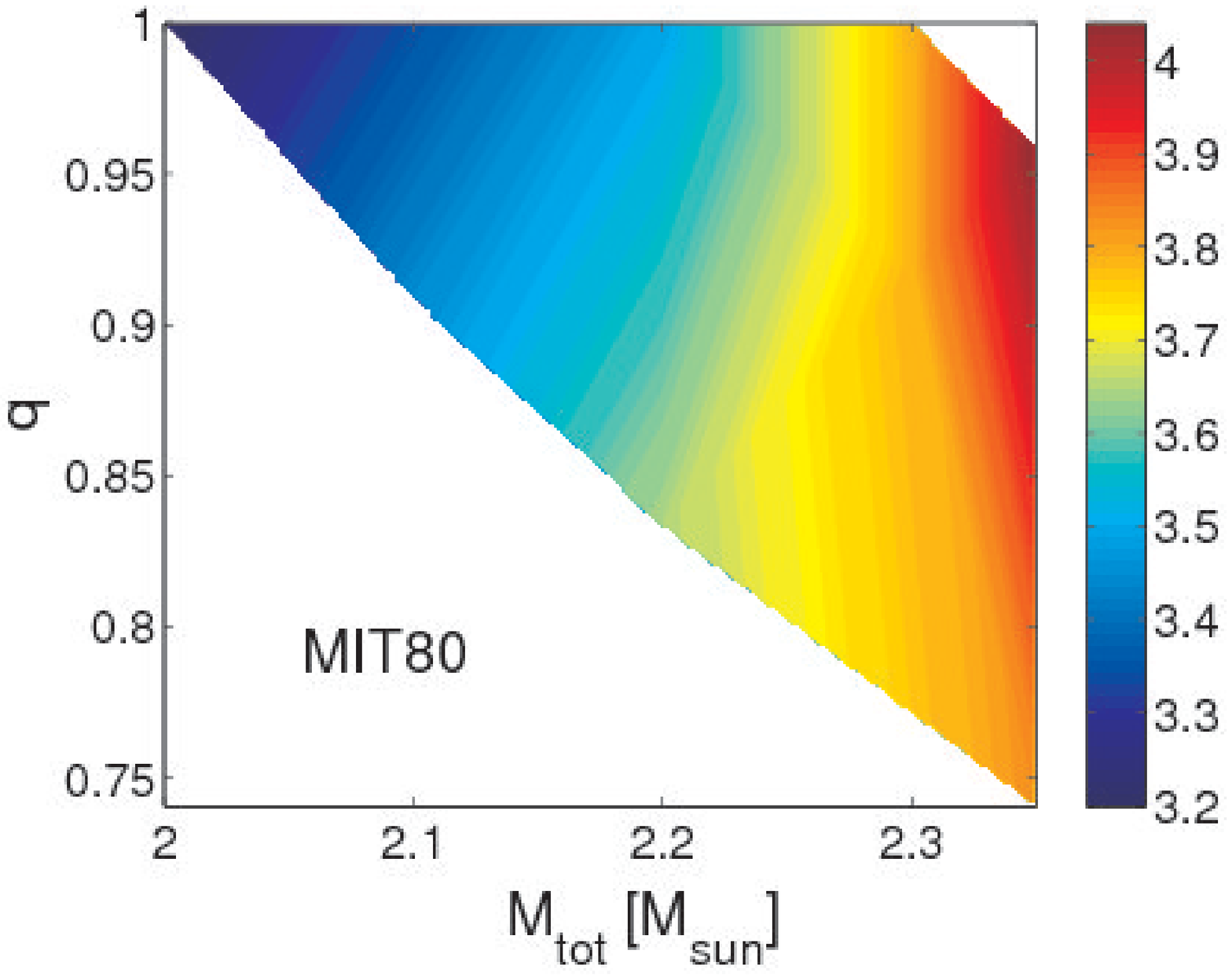}
\caption{Peak frequency of the postmerger ringdown color-coded in kHz as a function of the total binary mass $M_{\mathrm{tot}}$ and of the mass ratio $q$ for binary mergers with the MIT80 EoS.}
\label{fig:fpeak80}
\end{center}
\end{figure}

\begin{figure}
\begin{center}
\includegraphics[width=8.5cm]{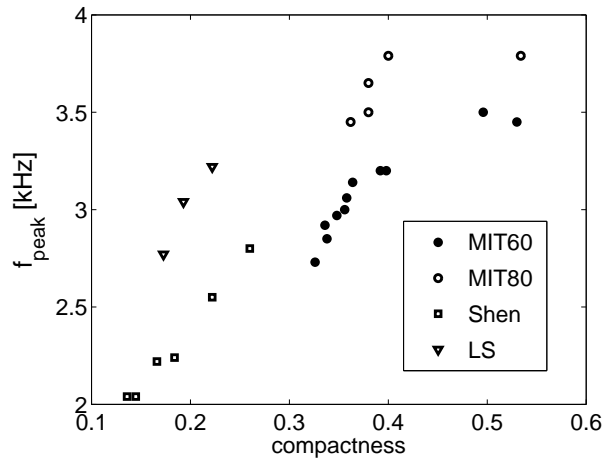}
\caption{ Peak frequencies of the postmerger ringdown as a function of the remnant compactness for all EoSs considered in this study. While the surface of SS merger remnants can be clearly determined, we define the size of NS merger remnants arbitrarily by the isodensity surface with $\rho=10^{12}\;\mathrm{g/cm^3}$. The mass of the remnant is approximated by the total mass $M_{\mathrm{tot}}=M_1+M_2$. The compactness is then given as $M_{\mathrm{tot}}$ divided by the remnant radius in the equatorial plane in isotropic coordinates at the end of the simulation when an axisymmetric hypermassive object has formed. Note that the differentially rotating remnant is highly deformed with an axis ratio of 2:2:1. For each EoS an approximately linear relation between $f_{\mathrm{peak}}$ and the compactness is obtained. The outliers are relatively massive configurations, which collapse to a BH during the simulation time.}
\label{fig:fpeakcomp}
\end{center}
\end{figure}

Figures~\ref{fig:fmax60},~\ref{fig:fmax80},~\ref{fig:fpeak60} and~\ref{fig:fpeak80} present in overview the characteristic frequencies for all models that we computed with the MIT60 and MIT80 EoSs. The results are displayed in dependence of the total binary mass $M_{\mathrm{tot}}=M_1+M_2$ and of the mass ratio $q=M_1/M_2$. Note that the peak frequency of the postmerger ringdown is given only for the configurations that form a hypermassive remnant.

For the maximum frequency of the inspiral we find a monotonic, but not exclusive, dependence on the total mass of the binary as shown in Figs.~\ref{fig:fmax60} and~\ref{fig:fmax80}. This trend is also present when considering $f_{\mathrm{max}}$ as a function of the compactness of the initial stars as given for all symmetric configurations in Fig.~\ref{fig:fmaxcomp}. The more compact the stars are, the higher the frequency is that is reached during the inspiral. Remarkably, $f_{\mathrm{max}}$ seems to follow an approximately linear function independent of the EoS. This confirms that the inner structure has a small effect on the inspiral phase. Only the relatively massive configurations describe a kink that deviates from the nearly linear behavior. The mass ratio also seems to have an influence, because in Figs.~\ref{fig:fmax60} and~\ref{fig:fmax80} $f_{\mathrm{max}}$ slightly decreases for lower values of $q$. Qualitatively the same result was found for the mergers of NSs (see \cite{2007PhRvL..99l1102O}). Our findings for the maximum GW frequency for the inspiral phase of SSs are also in agreement with \cite{GondekRosinska:2008nf}, where symmetric binaries were considered and a similar behavior was seen.

Figures~\ref{fig:fpeak60} and \ref{fig:fpeak80} reveal that for both EoSs the frequency of the postmerger peak increases monotonically with the total mass of the system. We interpret this dependence as a consequence of the growing compactness of the hypermassive remnant (see Fig.~\ref{fig:fpeakcomp}), which is determined by the total system mass. The compactness in turn determines the rotational period and the frequency of oscillations of the deformed object that produces the GW signal. Corresponding results for neutron mergers have been reported by the authors of Ref. \cite{2007PhRvL..99l1102O}, whose findings we confirm by our additional NS merger simulations.

In addition, one observes that the mass ratio hardly affects the value of $f_{\mathrm{peak}}$. As mentioned above, the dynamics of the postmerger remnant are determined by the total mass of the single object that forms in the merger and is therefore nearly independent of the binary mass ratio.

\subsection{Equation of state dependence and thermal effects}
\begin{figure}
\begin{center}
\includegraphics[width=8.5cm]{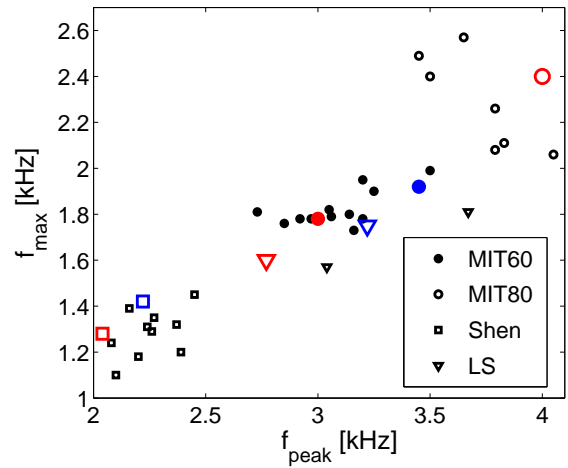}
\caption{Maximal frequencies during inspiral and peak frequencies of the ringdown of the postmerger remnant for all configurations that form a hypermassive object. Results for all EoSs used in this study are plotted. The blue symbols correspond to the binary configurations with two 1.35~$M_{\odot}$ stars, the red ones mark the frequencies for merger events of two stars with 1.2~$M_{\odot}$. Note that there is no blue symbol for the MIT80 EoS because the configuration with two 1.35~$M_{\odot}$ stars leads to an immediate BH formation. The compilation includes data from \cite{2007PhRvL..99l1102O}.}
\label{fig:eos}
\end{center}
\end{figure}

Obviously, a comparison of the influence of the different EoSs for SQM and for ordinary nuclear matter is of great importance. Figure~\ref{fig:eos} compiles the characteristic frequencies for the MIT60, MIT80, LS, and Shen models. The SS EoSs exhibit the tendency of being located in the upper right part of the plot, while LS and Shen give relatively low $f_{\mathrm{max}}$ between about 1.1 kHz and roughly 1.8 kHz for the maximum values. One should keep in mind that we did not compute the same sample of binary configurations for all EoSs, because the different EoSs support differently high masses and determine the system properties for which a hypermassive remnant forms and $f_{\mathrm{peak}}$ can be extracted. However, a comparison of the same binary setup for all EoSs is possible in the cases of the colored symbols in Fig.~\ref{fig:eos} (blue: 1.35$M_{\odot}$+1.35$M_{\odot}$ configuration; red: 1.2$M_{\odot}$+1.2$M_{\odot}$ configuration) and the trends visible are compatible with the rest of the data.

Again the diagram can be understood by means of the compactness of the merging stars and postmerger remnants: The more compact the stars or the merger remnants are, the higher are $f_{\mathrm{max}}$ and $f_{\mathrm{peak}}$, respectively (see Figs.~\ref{fig:fmaxcomp} and~\ref{fig:fpeakcomp} and compare also with Fig.~\ref{fig:MR}). Remarkably, MIT60 and LS show an overlap of $f_{\mathrm{max}}$ and $f_{\mathrm{peak}}$ values, which means that they can not be distinguished by a determination of these characteristic frequencies in GW measurements. 

In this context we refer to the discussion in \cite{2008arXiv0812.4248B}, which predicts an enhanced observable flux of strangelets in cosmic rays if the strange matter hypothesis was true and quark matter was described by an EoS similar to MIT60. In contrast, if the properties of SQM caused SSs to be as compact as those obtained with the MIT80 EoS, this should be clearly visible in the high characteristic frequencies of the GW signal. On the other hand, a nuclear EoS like the Shen EoS appears to be distinguishable from SQM because the MIT60 EoS represents a lower limit with respect to the compactness of SSs.

The existence of strangelets may be also crucial to discern self-bound SSs from hybrid stars, i.e. from NSs with cores of quark matter that is not absolutely stable and thus cannot form self-bound quark matter nuggets (see e.g.~\cite{1996csnp.book.....G}). The rich diversity of hybrid stars (see~\cite{2005ApJ...629..969A}) also includes objects with stellar parameters comparable to those obtained with the LS EoS or the MIT60 EoS. Therefore, hybrid stars might populate the ``LS-MIT60-area'' in Fig.~\ref{fig:eos}. However, for many hybrid star models we expect also the GW signals to be discriminating, since these objects can have a compactness, as a characteristic quantity determining the GW features, well below the one of NSs described by the LS EoS \cite{2005ApJ...629..969A}. This clearly distinguishes them from SSs. For an investigation of the differences of the GW signals of ordinary NSs and hybrid stars we refer the reader to \cite{2004MNRAS.349.1469O}.

\begin{figure}
\begin{center}
\includegraphics[width=8.5cm]{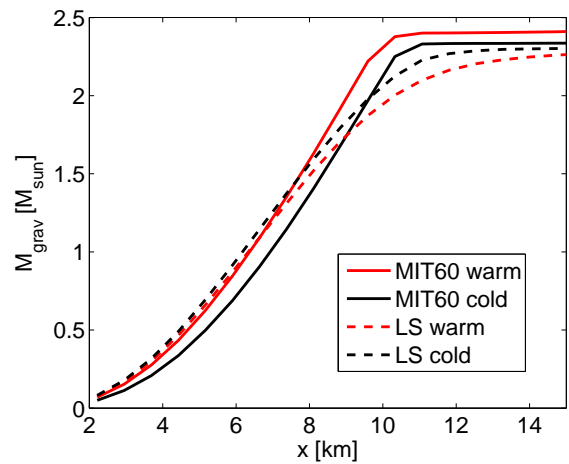}
\caption{Enclosed gravitational mass within an ellipsoid with the semiaxes $a=x$, $b=x$ and $c=x/2$ for the 1.2$M_{\odot}$+1.35$M_{\odot}$ binaries. Shown are results with the LS EoS and the MIT60 EoS including (red curves) and neglecting (black curves) thermal effects. Note that within general relativity the gravitational mass is only defined in isolation. Here $M_{\mathrm{grav}}$ denotes the enclosed contribution to the Arnowitt-Deser-Misner mass, neglecting the extrinsic curvature terms. Using the enclosed rest mass yields a similar result. The distance $x$ is given in isotropic coordinates.}
\label{fig:encmas}
\end{center}
\end{figure}

The comparison of the GW signals from cold and warm simulations confirms the findings of the previous section that the additional pressure support by nonzero temperature effects is important for NS mergers, while for the SS models the change of the gravitational mass in the zero temperature simulations is the most relevant aspect (see Table~\ref{tab:config}). As one might expect, $f_{\mathrm{max}}$ is very similar because thermal effects do not play a role during the inspiral. In fact, the variation remains below the relatively large uncertainties of about 0.1~kHz for $f_{\mathrm{max}}$ that are connected with the numerical determination of the merging time.

Comparing the ring-down signals, one finds that for the models with the LS EoS $f_{\mathrm{peak}}$ is higher in the cold than in the warm simulations (Table~\ref{tab:config}). Figure~\ref{fig:encmas} provides the explanation of this result. It displays the mass enclosed by ellipsoidal surfaces with varied distance from the center of the remnant. One can see that for the nuclear LS EoS the cold remnant is more compact because of the missing thermal pressure support. Consequently, the peak frequencies are higher.

In contrast to the NS models, we find lower peak frequencies in the cold SS merger simulations. As visible in Fig.~\ref{fig:encmas}, in the case of SSs the cold remnants are less compact due to the lower gravitational mass, while the remnant size is essentially unaltered since the thermal pressure support is not important for SQM. The reduced compactness explains the lower peak frequencies for SS merger simulations with $T=0$.

\subsection{Occurrence of a prompt collapse}\label{ssub:prompt}
The occurrence of a prompt collapse might be another feature that can be used to discern SSs from NSs. From GW measurements the chirp mass and most probably the individual masses of the binary components can be deduced, and therefore the total mass and the mass ratio. In addition, by analyzing the power at higher frequencies it should be possible to judge from a measurement whether a hypermassive postmerger remnant or a BH formed. This idea has already been brought forward in \cite{2005PhRvL..94t1101S}. Therefore we checked how the mass limit for a prompt collapse depends on the EoS. For the sake of clarity we consider only equal-mass binaries, but we expect the analysis to also hold for asymmetric systems.

For the LS EoS the minimum total mass $M_{\mathrm{collapse}}$ for
configurations that undergo a prompt collapse instead of forming a
hypermassive object is between 2.8~$M_{\odot}$ and 2.9~
$M_{\odot}$. The corresponding mass limit for the MIT60 EoS is
slightly higher than 2.7~$M_{\odot}$, but below 2.8~$M_{\odot}$. As mentioned above, the symmetric binary with $M_{\mathrm{tot}}=2.7~M_{\odot}$ does not collapse promptly but only $\sim 4$~ms after the hypermassive object has formed. Therefore, by this property a marginal discrimination of these two EoSs is possible. For MIT80 we find $M_{\mathrm{collapse}}\lesssim 2.4~M_{\odot}$, which distinguishes this EoS clearly from the majority of nuclear EoSs, because the LS EoS represents a fairly extreme case in terms of softness, NS compactness and $M_{\mathrm{collapse}}$. The threshold mass for the Shen EoS is about 3.5~$M_{\odot}$. For asymmetric systems we refer to Fig.~1 in \cite{2008arXiv0812.4248B}, which shows the borderline between binary configurations that result in a prompt collapse and those that lead to the formation of a hypermassive object.

The criterion discussed here can be considered as very simple and straightforward in its application to observational data. This makes it attractive for an analysis of merger signals and for the determination of basic constraints on the EoS. A drawback, however, may be the fact that GW measurements for a larger set of events will be needed to constrain the mass limit for the direct BH formation. Alternatively, the delay time between plunge, defined as the time when the GW amplitude becomes maximal, and collapse, which ultimately occurs for every hypermassive object, will be discussed as source of information in \cite{rezzolla}. For the models described in detail in Sect.~\ref{sec:sim}, we report the delay times $\tau_{\mathrm{delay}}$ in Table~\ref{tab:config} for the cases where the collapse occurs during the simulation. Otherwise the time until the end of the simulation gives an lower limit on $\tau_{\mathrm{delay}}$, but the true value may be significantly larger.

\subsection{Gravitational-wave luminosity}
As argued above it is not possible in all cases to distinguish NS mergers from SS mergers by extracting relatively simple characteristic frequencies from a GW measurement. If the mass-radius relations of compact stars were similar even only in the most relevant range of stellar masses, the frequencies $f_{\mathrm{max}}$ and $f_{\mathrm{peak}}$ may be in the same range of values for different EoSs. A more detailed analysis may yield additional discriminating information but will also require a higher quality of the signal determination.

Besides the method of matched filtering, which requires the knowledge of the signal searched for, analysis pipelines of GW detectors have the ability to measure excess power in time-frequency domains \cite{abbott-2009,abbott-2009-2}, which allows to search for unmodeled signals. Ratios of frequency-integrated energies in the luminosity spectrum, for example, can be considered as signal characteristics that can be extracted from the observational data without knowing the signals in detail. For instance one could consider the ratio between the energy $\Delta E_{\mathrm{pm}}$ radiated away by the postmerger remnant within the first 5~ms after merging and the energy $\Delta E_{\mathrm{in}}$ emitted in the last 3~ms of the inspiral before the coalescence. The time of merging is defined as the time when the GW amplitude becomes maximal. This quantity was introduced for characterizing binary NS merger GW signals in \cite{2007PhRvL..99l1102O}. The energies can be determined either by an integral over the luminosity spectrum or more easily by a time integral over the luminosity given directly by the quadrupole evolution $dE/dt=1/5 \langle \dddot Q_{\mathrm{ij}} \dddot Q_{\mathrm{ij}} \rangle$. Both ways of evaluation yield quantitative agreement within $~20\%$ when applied to the results of our numerical simulations. For simplicity we consider the latter expression.

\begin{figure}
\begin{center}
\includegraphics[width=8.5cm]{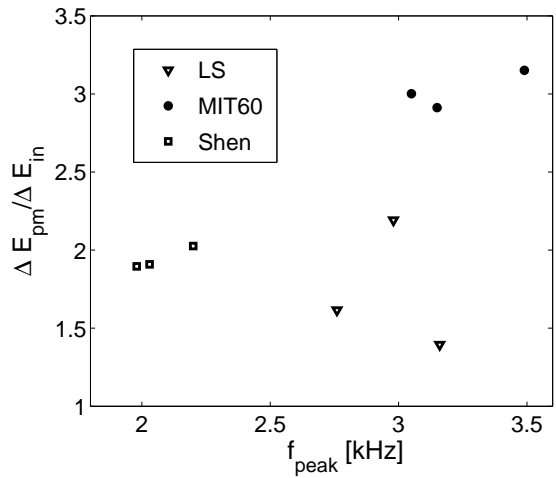}
\caption{Ratios of the GW energy emitted by the postmerger remnant within 5~ms after the coalescence to the energy radiated during the last 3~ms of the inspiral as a function of the peak GW frequency of the postmerger remnant. Shown are results for the binary configurations with 1.2~$M_{\odot}$ and 1.2~$M_{\odot}$ (left datapoints of each EoS cluster), with 1.2~$M_{\odot}$ and 1.35~$M_{\odot}$ (central datapoints), and with 1.35~$M_{\odot}$ and 1.35~$M_{\odot}$ (right datapoints) for the Shen, LS, and MIT60 EoS.}
\label{fig:lum}
\end{center}
\end{figure}

Figure~\ref{fig:lum} shows the ratio of $\Delta E_{\mathrm{pm}}$ to $\Delta E_{\mathrm{in}}$ as a function of the peak frequency $f_{\mathrm{peak}}$ of the postmerger ringdown for all possible binary configurations of stars with 1.2~$M_{\odot}$ and 1.35~$M_{\odot}$, and for the EoSs of Shen, LS and MIT60. Note that $f_{\mathrm{peak}}$ increases with the total mass of the system as shown in Figs.~\ref{fig:fpeak60} and \ref{fig:fpeak80}, allowing for an unambiguous identification of the three configurations in this plot (the least massive binaries are located to the left of the EoS-clusters of datapoints). Clearly one can see that the ratio $\Delta E_{\mathrm{pm}}/\Delta E_{\mathrm{in}}$ is a valuable measure to discriminate NS mergers described by the LS EoS from SS mergers with the MIT60 EoS. One may suspect such differences already from a comparison of the wave amplitudes during the postmerging phase plotted in Figs.~\ref{fig:gw12135} and~\ref{fig:gw135135}.

\begin{figure}
\begin{center}
\includegraphics[width=8.5cm]{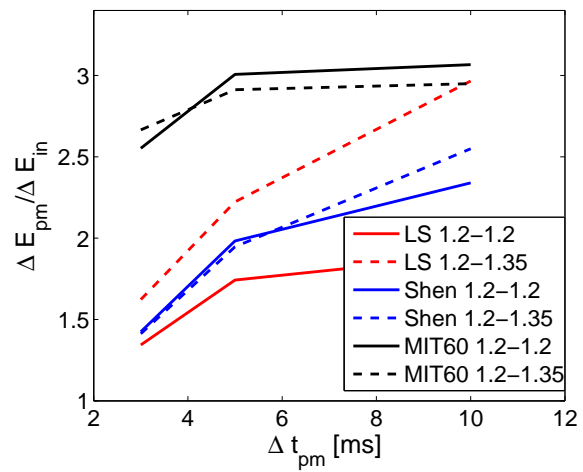}
\caption{Ratios $\Delta E_{\mathrm{pm}}/\Delta E_{\mathrm{in}}$ for different emission periods $\Delta t_{\mathrm{pm}}$ of GWs during the postmerging phase.}
\label{fig:lumevol}
\end{center}
\end{figure}

Motivated by the differences in the time evolution of the GW amplitudes visible in Figs.~\ref{fig:gw12135} and~\ref{fig:gw135135}, we also computed the energy ratios for postmerging time intervals $\Delta t_{\mathrm{pm}}$ of 3~ms, 5~ms, and 10~ms. Figure~\ref{fig:lumevol} reveals a noticeable discrepancy between SS mergers and coalescence events of ordinary NSs. Although the energy ratio is initially higher for SS postmerging remnants, the luminosity of these objects decays much faster and the value of $\Delta E_{\mathrm{pm}}/\Delta E_{\mathrm{in}}$ saturates after about 5~ms. In contrast, this ratio increases for NS mergers over emission times even longer than 5~ms after the coalescence. Therefore, also the decay time of the postmerging emission should be considered as a feature that can be used to discern the GW signal of SS mergers and NS mergers, in particular in cases when the characteristic signal frequencies $f_{\mathrm{max}}$ and $f_{\mathrm{peak}}$ do not allow for an unambiguous discrimination.

As visible in Figs.~\ref{fig:spec12135} and~\ref{fig:spec135135}, the GW luminosity spectra of NS mergers employing the LS EoS show a prominent gap at frequencies slightly lower than $f_{\mathrm{peak}}$. This feature occurs in all of our models with the LS EoS, while it is less pronounced in the cases of mergers with the MIT60 EoS. The gap is also visible in the Fourier transformed waveform and it occurs for each emission direction. This suggests the use of the gap to break the degeneracy of the LS and MIT60 EoSs in $f_{\mathrm{max}}$ and $f_{\mathrm{peak}}$ (Fig.~\ref{fig:eos}); we focus on these two EoSs because the characteristic frequencies do not allow their discrimination as in the cases of the other EoSs.

\begin{figure}
\begin{center}
\includegraphics[width=8.5cm]{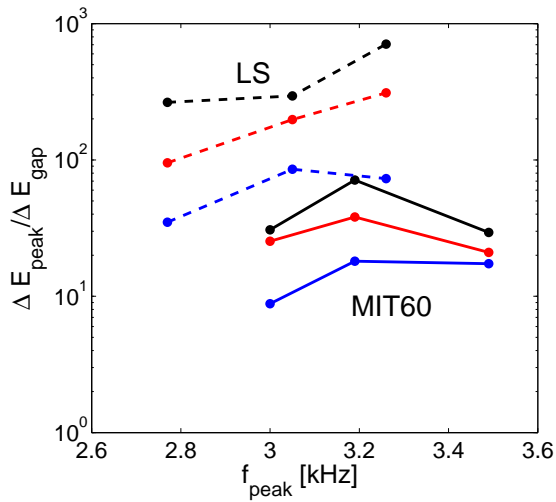}
\caption{Ratios $\Delta E_{\mathrm{peak}}/\Delta E_{\mathrm{gap}}$ of the GW energy emitted in a frequency window $\Delta f$ around the peak frequency to the energy emitted in a frequency window $\Delta f$ around the gap frequency as a function of the peak frequencies of the postmerger ringdown. The colors correspond to differently chosen frequency windows $\Delta f$ (see main text), therefore lines of the same color are to be compared (blue: 250 Hz; red: 100 Hz; black: 50 Hz). Note the logarithmic scale of the ratio. Shown are results for the binary configurations with 1.2~$M_{\odot}$ and 1.2~$M_{\odot}$ (left datapoints), with 1.2~$M_{\odot}$ and 1.35~$M_{\odot}$ (central datapoints), and with 1.35~$M_{\odot}$ and 1.35~$M_{\odot}$ (right datapoints) for the LS EoS (dotted lines) and MIT60 EoS (solid lines). }
\label{fig:gap}
\end{center}
\end{figure}

We introduce frequency intervals with a given width $\Delta f$ placed centered around $f_{\mathrm{peak}}$ and $f_{\mathrm{gap}}$. The latter frequency is defined by the position of the local minimum that defines the gap in the luminosity spectrum (Figs.~\ref{fig:spec12135} and~\ref{fig:spec135135}). The values of $f_{\mathrm{gap}}$ for the more massive binary configurations are listed in Table~\ref{tab:config}, while for the 1.2$M_{\odot}$+1.2$M_{\odot}$ binary $f_{\mathrm{gap}}$ equals 2.37 kHz for the LS EoS and 2.57 kHz in the case of the MIT60 EoS. Integrating the luminosity spectra in these frequency windows and comparing the energies provides a characteristic quantity for a quantitative description of the gap feature of NS merger models with the LS EoS. This energy ratio $\Delta E_{\mathrm{peak}}/\Delta E_{\mathrm{gap}}$ may be deduced easily from observational data by analyzing the excess power in different frequency domains \cite{abbott-2009,abbott-2009-2}. Figure~\ref{fig:gap} shows the ratio $\Delta E_{\mathrm{peak}}/\Delta E_{\mathrm{gap}}$ of binary mergers with the LS and MIT60 EoSs for different choices of $\Delta f$. Equal colors correspond to the same value of $\Delta f$. For smaller frequency windows the difference in the ratios becomes more enhanced as one expects from Figs.~\ref{fig:spec12135} and~\ref{fig:spec135135}. One can clearly see that for all choices of $\Delta f$ the ratio $\Delta E_{\mathrm{peak}}/\Delta E_{\mathrm{gap}}$ allows to distinguish the GW emission of mergers described by the LS EoS from the signal of the SS coalescence events with the MIT60 EoS.

\section{Summary and conclusions}\label{sec:con}

We have reported on simulations of SS mergers and discussed our results in comparison to NS coalescence events. In particular, we focussed on the GW signals and their potential to provide information about the existence of SSs.

We found that the dynamical behavior of SS mergers is fundamentally different, which can be understood by the higher compactness of SSs and their merger remnants, the self-binding of SQM and the different influence of thermal effects. While NS mergers form a dilute halo or toruslike structure around a dense, hypermassive, differentially rotating remnant, the remnant of merging SSs is bounded by a sharp surface as were the initial stars. Only by the formation of thin tidal arms relatively late in the evolution of the remnant, matter gets shed off the central object and forms a fragmented thin disk in the equatorial plane.

In order to estimate the importance of thermal effects during NS and SS mergers we compared our models with simulations where zero temperature was imposed by extracting the thermal energy as most efficient cooling would do. For both kinds of stars we observed a sensitive dependence of the dynamics of the system on thermal effects, affecting for instance the development and structure of the outer remnant parts of NS mergers. While the estimated relic torus mass after collapse of the remnant to a BH is in general higher if thermal effects were neglected, a nonzero temperature of NS matter leads to an inflated, dilute halolike torus in contrast to a much thinner, more disklike structure for $T=0$. For SS mergers we found a shorter time delay for the BH formation, whereas in the case of NSs this time interval is stretched as reported in \cite{2008PhRvD..78h4033B}. This difference can be understood as a consequence of the additional gravitating effect of the thermal energy, which in the case of very dense SSs is not overcompensated by thermal pressure effects (different from the NS case).

The analysis of the GW signals emitted by SS mergers revealed that already by means of relatively simple characteristic features of the signal it may be possible to decide whether SS or NS mergers produced the emission. In particular, we found that the maximal frequency during the inspiral and the frequency of the ringdown of the postmerging remnant are in general higher in the case of SS mergers in comparison to NS mergers. Whether this criterion can be used to finally decide on the strange matter hypothesis depends on the particular stellar properties associated with the EoS of high-density matter. For a similar mass-radius relation within a certain mass range, meaning relatively compact NSs or less compact SSs (the LS-MIT60 scenario), the determination of these frequencies might not be decisive. However, taking additional characteristics of the GW luminosity into consideration will allow for a discrimination. In this context we discussed the occurrence of a prompt collapse, the ratios of the GW energy emitted in the postmerger phase to the energy radiated away during the inspiral phase, the growth rate of the energy emission associated with the postmerger ringdown signal, and the appearance of a gap in the GW luminosity spectrum. In addition, it was shown in \cite{2008arXiv0812.4248B} that cosmic ray experiments will yield information about the strange matter hypothesis and may clarify the situation, especially in the case when GW signals are not conclusive as in the LS-MIT60 scenario. Therefore we expect that despite the rather uncertain event rate the upcoming advanced GW detectors LIGO and VIRGO may provide valuable data to decide about this long-standing question, and it appears likely that one will thus gain fundamental information about the properties of high-density matter. Moreover, future GW detectors like the Einstein telescope \cite{et} and the DUAL detector \cite{dual} will have a higher sensitivity in the high-frequency domain above 1~kHz, where characteristic features occur. Signals measured with these future instruments will therefore allow for a more detailed analysis.

Finally, we would like to point out that also other forms of self-bound matter have been discussed in the literature (see e.g.~\cite{1977JETPL..25..465V,1990PhR...192..179M,2007ASSL..326.....H}). For instance pion condensates may lead to stellar objects similar to SSs, and also abnormal nuclei similar to strangelets may exist in this case. Therefore, the study presented in this work should be generalized to these states of matter, where quantitative considerations require the knowledge of the specific EoS. However, for stellar properties comparable to those of SSs we expect qualitatively the same behavior and the same GW features as for SSs. From this discussion it is clear that only a multimessenger approach can bring a final answer. In addition to GW measurements and cosmic ray experiments, this may in particular include the observation of isolated compact stars to derive the cooling history of these objects, which is significantly different for NSs, SSs and self-bound pion-condensed stars (see~\cite{1999paln.conf.....W,2000ApJ...533..406B,2004ARA&A..42..169Y,2005PhRvC..71d5801G}).

Several improvements and extensions may supplement this first study of SS mergers in the future. Besides refinements in the methodical approach like a fully relativistic treatment and the inclusion of magnetic fields (only important during the merging phase if very strong, but must be expected to grow afterwards and affect the ringdown phase) and radiative effects, one would also like to account for the effects of quark interaction and color superconductivity, employ models beyond the MIT bag model, and discuss other forms of self-bound matter. Moreover, a determination of the modes excited in the merger remnants may be a promising way to develop a better understanding of the origin of the characteristic properties of the GW spectra. Furthermore a study of SS-BH mergers will yield further insights about the behavior of such systems in contrast to NS-BH binaries.

\begin{acknowledgments}
It is a pleasure to thank G.~Pagliara, I.~Sagert, and J.~Schaffner-Bielich for providing the strange quark matter equations of state and for helpful discussions. This work was supported by the Sonderforschungsbereich Transregio 7 ``Gravitational Wave Astronomy", by the Sonderforschungsbereich Transregio 27 ``Neutrinos and Beyond", and the Cluster of Excellence EXC 153 ``Origin and Structure of the Universe" of the Deutsche Forschungsgemeinschaft, and by CompStar - a research networking program of the European Science Foundation. The computations were performed at the Rechenzentrum Garching of the Max-Planck-Gesellschaft and at the Leibniz-Rechenzentrum Garching.
\end{acknowledgments}

\bibliography{references}

\end{document}